\documentclass[astrosymb, twocolumn, twocolappendix]{aastex631}
\usepackage{enumitem}
\usepackage{amsmath}
\usepackage{listings}

\shorttitle{KBonus: Kepler Background Sources}
\shortauthors{Mart\'inez-Palomera et al.}
\graphicspath{{./}{figures/}}

\begin{document}

\title{Kepler Bonus: Light Curves of Kepler Background Sources}

\correspondingauthor{Jorge Martínez-Palomera}
\email{palomera@baeri.org, jorgemarpa@ug.uchile.cl}

\author[0000-0002-7395-4935]{Jorge Martínez-Palomera}
\affiliation{Bay Area Environmental Research Institute, P.O. Box 25, Moffett Field, CA 94035, USA.}
\affiliation{NASA Ames Research Center, Moffett Field, CA, USA}

\author[0000-0002-3385-8391]{Christina Hedges}
\affiliation{NASA Goddard Space Flight Center, Greenbelt, Maryland, United States}
\affiliation{University of Maryland, Baltimore County, 1000 Hilltop Circle, Baltimore, Maryland, United States}

\author[0000-0003-4206-5649]{Jessie Dotson}
\affiliation{NASA Ames Research Center, Moffett Field, CA, USA}


\begin{abstract}

NASA's \textit{Kepler} primary mission observed about 116 $deg^2$ in the sky for 3.5 consecutive years to discover Earth-like exoplanets.
This mission recorded pixel cutouts, known as Target Pixel Files (TPFs), of over $200,000$ targets selected to maximize the scientific yield.
The Kepler pipeline performed aperture photometry for these primary targets to create light curves.
However, hundreds of thousands of background sources were recorded in the TPFs and have never been systematically analyzed.
This work uses the Linearized Field Deblending (LFD) method, a Point Spread Function (PSF) photometry algorithm, to extract light curves.
We use Gaia DR3 as input catalog to extract $606,900$ light curves from long-cadence TPFs.
$406,548$ are new light curves of background sources, while the rest are Kepler's targets.
These light curves have comparable quality as those computed by the Kepler pipeline, with CDPP values $<100$ ppm for sources $G<16$.
The light curve files are available as high-level science products at MAST.
Files include PSF and aperture photometry, and extraction metrics.
Additionally, we improve the background and PSF modeling in the LFD method.
The LFD method is implemented in the \texttt{Python} library \texttt{psfmachine}.
We demonstrate the advantages of this new dataset with two examples; deblending of contaminated false positive Kepler Object of Interest identifying the origin of the transit signal; and the changes in estimated transit depth of planets using PSF photometry which improves dilution when compared to aperture photometry.
This new nearly unbiased catalog enables further studies in planet search, occurrence rates, and other time-domain studies.

\end{abstract}


\section{Introduction} \label{sec:intro}

NASA's \textit{Kepler} mission delivered to the community one of the most precise time series datasets ever produced.
During its primary mission, \textit{Kepler} observed more than $200,000$ target stars \citep{2010Sci...327..977B}. 
\textit{Kepler} found more than $2,600$ exoplanet candidates \citep{Thompson_2018}, observed numerous supernovae from earliest stages of explosion \citep{2015Natur.521..332O, 2016ApJ...820...23G, 2019ApJ...870...12L}, and more than $2,900$ eclipsing binary systems \citep{2016AJ....151...68K}. The Kepler mission had a significant impact on a range of astrophysical domains owing to its precise, accurate time series of a large sample of stars, for the 3.5 year prime mission. 
Yet its contribution to time domain astronomy is not finished. Thanks to the use of current catalogs and methods it is possible to significantly expand the volume of data products originating from Kepler's observations. This work presents a catalog of $606,900$ light curves including $406,548$ from new sources and $200,352$ Kepler targets.

\subsection{\emph{Kepler} Target Selection Function}

Kepler's primary mission selected over 200,000 targets to maximize the yield of Earth-like exoplanet discoveries \citep{2010ApJ...713L.109B}. 
These targets were selected from approximately half a million stars {brighter than 16th magnitude in the Kepler passband ($K_{p}$). 
The selection used stellar parameters to estimate the radius of the smallest planet detectable in the habitable zone, the number of detectable transits, and samples per transit. 
These combined with a crowding metric for the photometric aperture and the target brightness resulted in a prioritization criteria that was used to rank and select the target list.
The target list mainly focuses on main-sequence G-type stars (half of the target sample) with a large fraction of them brighter than magnitude $K_{p} < 14$. 
The target list also includes M-type dwarfs and a small sample of hot main-sequence O- and B-type stars.
Using Gaia DR2 catalog \cite{2021AJ....161..231W} found that Kepler's target selection is nearly complete for main-sequence stars brighter than $K_{p} < 14$ mag and it is biased against binary systems. 
The study found that Kepler's selection favored cool dwarfs fainter at the faint end.
Additionally, the target selection effectively separated red giants from red dwarfs. 
The abovementioned study found a significant drop in the observed fraction of red giants at fainter magnitudes, particularly for low-luminosity, cool giants.
The same work also found no significant bias in target kinematics.

\subsection{\emph{Kepler} Data Products}

Kepler data products are available in three categories, discussed below; Target Pixel Files (TPFs), Light Curve Files (LCFs) and Full Frame Images (FFIs).

During its primary mission, \textit{Kepler} observed seventeen 93-days periods named quarters. 
The Kepler instrument consisted of 21 science modules, each having 4 output channels, for a total of 84 CCD channels.
The telescope rotated $90^{\circ}$ every quarter which led to the same objects being observed in the same CCD channel every 4 quarters.
The \emph{Kepler} telescope observed an approximately 116 squared-degree region of the sky at a cadence of 30 minutes.  
During the prime mission, pre-defined targets were downloaded as images and were converted to flux time-series on the ground. 
Target cutouts were centered on stars selected from the Kepler Input Catalog \citep[KIC, ][]{2011AJ....142..112B} and are typically 4 to 9 pixels around the target.
The Kepler Science Data Processing Pipeline \citep{Jenkins_2010}, produced two science products from these cutouts, the Target Pixel Files (TPFs) and the Light Curve Files (LCFs). 

TPFs contain the time series at the pixel level and the aperture mask used to compute the photometry of the target.
LCFs are flux time series of the target.
Both data products were created for a short 1-minute and a long 30-minute cadence mode. 
Short cadence targets required more onboard storage and different processing on the ground, and so were used on high-value targets only. 
All short cadence targets also produced long cadence products. 
In this work, we will only consider the long cadence targets, as these are available for the full Kepler sample. 
We leave any discussion or treatment of short cadence targets to future work.

\textit{Kepler}'s prime mission also downlinked single Full Frame Images (FFIs) of the entire field of view each month. FFIs were downloaded for calibration and diagnostic purposes \citep{2016ksci.rept....1V}. FFIs have an exposure time of 30 minutes but were only captured each month, then in this work, we will not use them for time series. We leave any discussion of the benefits of FFI data for extracting time series for future work. 

\emph{Kepler} light curves were extracted using Simple Aperture Photometry (SAP) with a pre-computed aperture mask. 
This aperture mask balanced the precision of the flux measurement while keeping the contamination from neighbors low. 
The LCFs also contain a corrected version of the SAP flux, the Presearch Data Conditioned Simple Aperture Photometry \citep[PDCSAP, ][]{2012PASP..124.1000S} (PDCSAP), which corrects for the systematics of the instrument. 
PDCSAP light curves are corrected using vectors of common trends from targets on the same detector channel, and largely address the systematics introduced by effects such as differential velocity aberration, and any spacecraft motion. 
Thanks to the instrument design, observation strategy, and data analysis the pipeline delivered light curves with high precision, enabling the detection of transits with $< 10$ ppm depth.

\subsection{Improving Kepler Light Curves with Gaia and PSF Photometry}

The \emph{Kepler} spacecraft was launched in 2009, after years of development. As such, the Kepler Input Catalog \citep[KIC, ][]{2011AJ....142..112B} predates the advent of the Gaia mission \citep{2016A&A...595A...1G}, and was assembled from earlier, less accurate catalogs. 

Using the KIC, the \textit{Kepler} pipeline performed photometry and computed optimized apertures for every target source, providing metrics that characterize the completeness of the flux and amount of contamination within the aperture. However, with updated knowledge from the Gaia catalog, we can now revisit these apertures and understand that many are significantly contaminated with background fainter sources ($G > 16$) or by bright neighbor sources.


In total, the \textit{Kepler} pipeline produced light curves for more than $206,000$ sources. 
But current more complete catalogs such as Gaia Data Release 3 \citep[Gaia DR3,][]{2022arXiv220800211G} lists a more than 1.4 million sources brighter than magnitude $G=19$ around the pixel cutouts.

In this work, we revisit \textit{Kepler}'s archival data to create a complete catalog of light curves using robust photometry, with our updated knowledge from Gaia. We use the Linearized Field Deblending (LFD) photometry method \citep{2021AJ....162..107H} to create light curves of $606,900$ sources. 
Of this, more than $400,000$ corresponds to newly extracted light curves of background sources, which doubles the number of \textit{Kepler} targets.
The LFD method provides a fast yet robust approach to perform Point Spread Function (PSF) photometry in \textit{Kepler}-like data.
LFD models the image at the pixel level to create a PSF model of the sources in the scene.
Here, the scene is defined as the collection of sources observed in a list of neighboring TPFs (here and after also called a stack of TPFs). 
The LFD method introduces perturbations to a mean PSF model in order to correct instrumental signals such as spacecraft motion and optic changes. 
Both the PSF fitting and evaluation are modeled as a linear problem and solved using least-square minimization. 
Through this, the LFD method is able to quickly estimate the PSF shape and perform PSF photometry.

The use of PSF photometry and current Gaia catalogs led to three main improvements over the original Kepler light curve catalogs. 
First, PSF photometry enables robust flux estimation and target deblending which is extremely relevant in crowded regions. 
These regions could be particularly problematic for aperture photometry due to close proximity of sources in the image and a varying range of source brightness contrast (the difference in magnitude between two nearby sources).
Secondly, Gaia catalogs provide precise astrometry and an improved census of objects in the field when compared to the KIC which enables access to a larger volume of sources. 
Thirdly, a blind massive light curve extraction leads to a nearly unbiased catalog useful for a better characterization of planet occurrence rate as well as further time-domain studies.

Here we present Kepler Bonus (KBonus), a catalog of extracted light curves that includes Kepler Targets and background sources.
All the light curves produced in this work are publicly available to the community as FITS Light Curve Files. 
These can be accessed via the Mikulski Archive for Space Telescopes (MAST) archive \footnote{KBonus Kepler Background \dataset[10.17909/7jbr-w430]{\doi{10.17909/7jbr-w430}}}.
We introduce new functionalities to the original LFD method to improve the PSF modeling and correction. 
These are available in version 1.1.4 of the \texttt{Python} package \texttt{psfmachine}\footnote{\texttt{psfmachine} v1.1.4 \url{https://github.com/SSDataLab/psfmachine/tree/v1.1.4}}. 
Additionally, accompanying this article we publish the KBonus repository\footnote{\url{https://github.com/jorgemarpa/KBonus/tree/main}} that shows examples of the processing pipeline and configuration files used for this work as well as an example of how to load the light curve files and its content.

This article is structured as follows. 
Section \ref{sec:data} details the characteristics of the data used for this work as well as the steps followed to compute the PSF models, photometry, flux metrics, and light curves.
In Section \ref{sec:results} we present our results: we characterize the quality of the extracted light curves, discuss the demographics of the resulting catalog, and showcase two science results using these light curves. 
In Section \ref{sec:limitations} we discuss the limitations of this work and in Section \ref{sec:future_work} the opportunities that this new unexplored dataset provides to the community.
Finally, Section \ref{sec:summary} summarizes this work.


\section{Data Processing} \label{sec:data}

We process the \emph{Kepler} data using the Python package \texttt{psfmachine} that performs Linearized Field Deblending (LFD) photometry \cite{2021AJ....162..107H}, a newly introduced type of rapid PSF photometry. 
In this work, we further improve \texttt{psfmachine} by adding a background estimator to remove rolling band noise, PSF models estimated with Kepler's FFIs, and the use of custom basis vectors to correct the scene motion due to differential velocity aberration. 
In this section, we describe the data used for this work, the additional analysis introduced from the original LFD work as well as the new algorithms and modules added to \texttt{psfmachine}. For an in-depth explanation of how the photometry of each source is extracted, we direct the reader to \cite{2021AJ....162..107H}.

\subsection{Kepler's Target Pixel Files} \label{subsec:tpfs}

Kepler observations are split in 
The Kepler pipeline delivered the observed data in the form of Target Pixel Files, a stack of pixels around each selected target for all observed cadences. 
We accessed a total of $204,933$ TPFs from MAST archive \footnote{\url{https://archive.stsci.edu/missions-and-data/kepler/kepler-bulk-downloads}} as well as other relevant engineering data (see Section \ref{subsec:bkg_model}).
Kepler 17 observing quarters and the 84 output channels distributed across the focal plane provide a natural strategy to process the TPFs to isolate instrument systematics spatially and temporally.
Therefore, we process the TPFs on a per-quarter-channel basis.
Within each quarter/channel combination, we split the list of available TPFs into ``batches'', in order to make the model fit memory efficient.

Each ``batch'' contains around 200 TPFs spatially sorted (i.e. 200 TPFs that are close on the detector). 
The batch size is not fixed due to the non-homogeneous distribution of targets around the focal plane and the changing total number of targets observed across quarters. 
We found that using $\sim 5\,000$ pixel time-series and $\sim 400$ unique sources, which is typically reached with $\sim 200$ TPFs, provides a robust fit of our mean and perturbed PSF model (see Sections \ref{subsec:PSF_model} and \ref{subsec:per_model}).  
In some crowded regions like around the open clusters NGC 6819 and NGC 6791, fewer TPFs are needed to constrain the model, owing to the source density in these clusters.

\subsection{Source Catalog} \label{subsec:gaia_cat}

The LFD method works by allowing the ``scene'' of stars to move as one, and each source to vary in brightness, but does not allow any individual source to move with respect to the others. The LFD method requires an astrometric catalog as input to fix the location of sources in the scene and to have a flux reference for each object. 
For this purpose, we use the Gaia DR 3 catalog.
Gaia DR 3 provides a complete catalog between magnitudes $G=12$ and $G=17$. It offers an astrometric precision of $0.4$ mas and a photometric precision of $6$ mmag at magnitude $G=20$ \citep{2023A&A...674A..32B}.  
We query the Gaia DR 3 catalog with the center of each available TPF, a generous search radius of the cutout size plus $16 \arcsec$ ($\approx$4 \emph{Kepler} pixels) to allow sources off the TPF edge and a magnitude limit of $G=19$. 
We propagate Gaia proper motions for every quarter observed by \textit{Kepler}.
We obtain a list of 1.4 million sources which acts as the input catalog for this work. 

To increase efficiency, we perform a more conservative query to the input catalog using the \texttt{psfmachine} API. 
We allow sources up to $4 \arcsec$ away from the TPF edge, remove sources brighter than $G=10$ to avoid saturated pixels and the nonlinear response of the CCD, and filter blended sources within $1 \arcsec$ by keeping the brighter objects. 
Highly blended sources, closer than $1\arcsec$ are difficult to successfully deblend, and imposing this filter helps to diminish the number of degenerated solutions.
The resulting catalog of successfully extracted sources contains $606,900$ entries. 
From the total, $200,352$ corresponds to Kepler targets for which the Kepler pipeline produced light curve files. 
The remaining $406,548$ objects correspond to background sources for which this work releases new light curves. 
Additionally, we perform a cross-match between the KIC and Gaia DR3 with a $2 \arcsec$ radius and accounting for proper motion, to identify original Kepler targets.

\begin{figure}[htb!]
    \centering
    \epsscale{1.15}
    \plotone{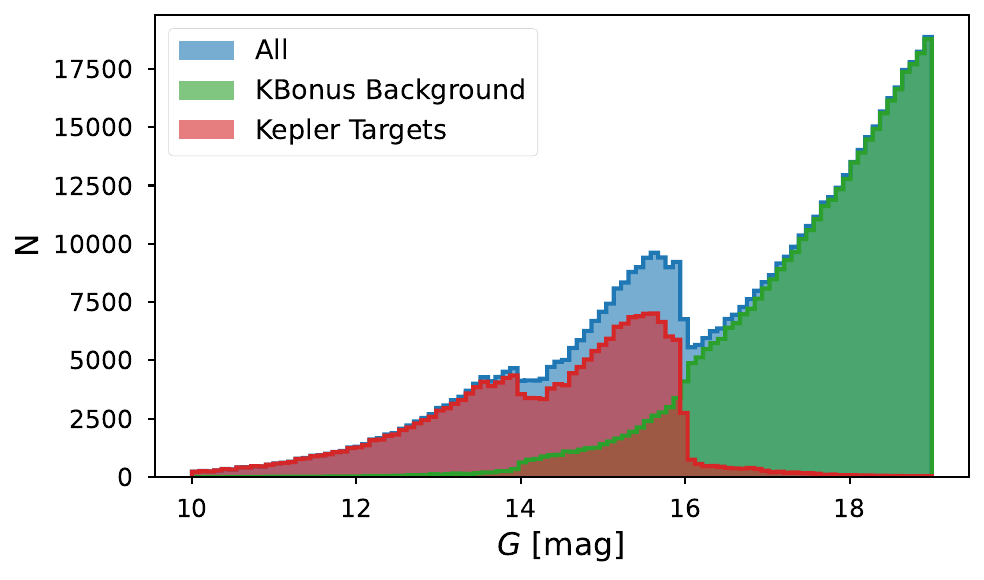}
    \caption{
    G-band magnitude distribution of Kepler targets (red), KBonus background sources (green), and both (blue).
    Kepler targets are the stars defined by the primary mission. These are in the center of the TPF and the Kepler pipeline created light curves for each of them.
    KBonus Background sources are stars that are fully or partially contained within a Kepler TPF but are not primary targets and therefore the Kepler pipeline did not analyze them. 
    }
    \label{fig:mag_dist}
\end{figure}

The apparent magnitude distribution of Kepler targets (Figure \ref{fig:mag_dist}) shows evidence of the target selection from the prime mission. This is reflected in, for example, the number cutoff at $G=16$ and the over-density around $G=13.8$ due to Sun-like stars being targeted. The apparent magnitude distribution of background sources shows no signs of selection bias based on star properties.

\begin{figure*}[htb!]
    \centering
    \epsscale{1.15}
    \plotone{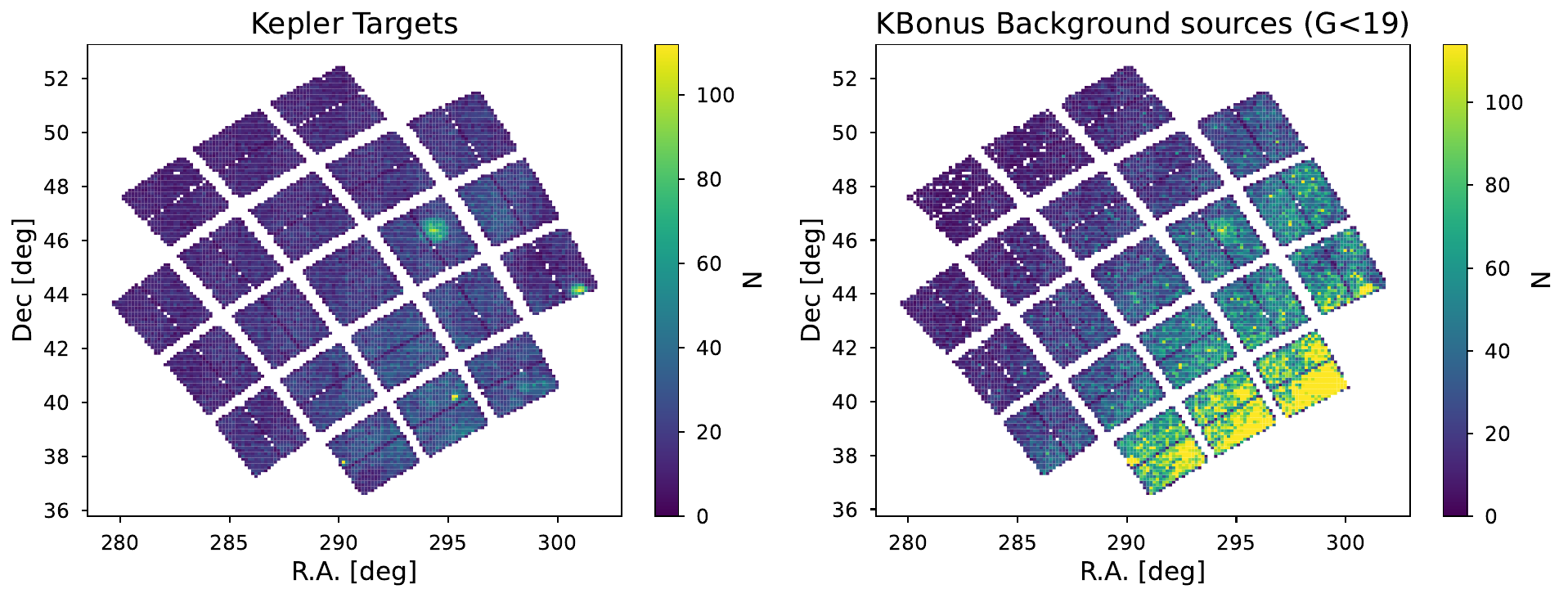}
    \caption{
    Spatial distribution (2D histogram) of Kepler targets (left) and KBonus background sources (right) brighter than magnitude 19th in the G-band across the Kepler field of view.  
    KBonus Background sources are stars that are fully or partially contained within a Kepler TPF but are not primary targets and therefore the Kepler pipeline did not analyze them. 
    The over densities on the left correspond to particular stellar clusters, while on the right the over densities closely follow the true underlying distribution of stars.
    }
    \label{fig:kepler_fov_density}
\end{figure*}

Figure \ref{fig:kepler_fov_density} shows the spatial distribution of Kepler targets and background sources across the field of view. 
The two high-density regions in the Kepler targets are the NGC 6819 and NGC 6791 open clusters. 
In contrast, the density of background sources shows an increasing number count closer to galactic latitudes. 

We removed saturated pixels and bleed columns from the sample to avoid introducing uninformative data points to the fitting process.
We used a conservative flux value to flag saturated pixels of $1.2e5 e^{-}/s$ and masked out up to three pixels around saturated ones to account for bleeding.
Additionally, we masked out pixels within a $800\arcsec$ from extremely bright sources ($G\leq8$) which typically exhibit halos due to internal reflections within the telescope. 
Sources that fall in these removed pixels are also ignored from the analysis. 

\subsection{Background Model} \label{subsec:bkg_model}

Kepler data shows a moving background signal known as ``rolling band'' \citep[Kepler Instrument Handbook,][]{2016ksci.rept....1V}. 
This correlated signal is more likely to occur on certain channels, at certain times of the year due to changes in the thermal background, and is difficult to model or predict.
The rolling band is observed as a shift in background level that moves almost parallel to the x-axis of the sensor. 
This artifact signal is small in amplitude, $\sim 20$ counts per pixel, but coadds to a large signal for large aperture photometry, and can adversely affect quiet sources. Crucially, this background is an additive signal and so can not be effectively removed by methods that divide out systematics (e.g. the CBV method.) 
The pipeline processed TPFs contain background-subtracted flux values as well as the subtract model computed by the \textit{Kepler} pipeline. 
Although the pipeline provides a good estimate of the background model, the \textit{Kepler} pipeline only addressed the rolling band issue by including a data quality flag \citep{2020ksci.rept....4C}.

In order to model and remove this signal, we build a background model as a function of time and pixel rows. 

Our method relies upon the strong row dependency of the rolling band signal to simplify the model and assume there is no signal in the orthogonal column direction. 
To constrain the model, we identify and model "background" pixels in the data set.

To identify ``background'' pixels, we use the source mask computed by \texttt{psfmachine} to find the pixels without a nearby source \citep[see section 4.2 in ][]{2021AJ....162..107H}, and we perform a sigma clipping to reject pixels that show significant variability. 
In addition, we augment the TPF pixel dataset with the mission background pixel data. 
The mission background data was taken during every quarter across every channel on a predefined grid distributed across each CCD  \citep{2016ksci.rept....1V}. 
Adding this dataset improve significantly the background model, especially in crowded regions where the TPF background pixel count is low.
We take the median average of the pixels in the column direction to find the average time series at every unique observed pixel row. 

We model the time series of the background pixels as two third-degree b-spline functions in both time ($t$) and pixel row number ($y$). We use knot spacings for the spline functions of 2 hours in the time direction, and 6 pixels in the row direction. 
This enables us to produce a flexible model that adapts to the fast-changing rolling band signal. This effectively builds a smooth model that averages values of pixels close in time and space.

We model the background of a batch of TPFs that have $n_{tot}$ total pixels data, $n_{bkg}$ of which are background pixels, and $l$ cadences as follows.
We build a design matrix $\mathbf{X}_{bkg}$ using the combination of two spline functions in time ($t$) and pixel row positions of the background pixel time-series ($y_{bkg}$):

\begin{equation}\label{eq:bkg_dm}
        \mathbf{X}_{bkg} = vec\left(\begin{bmatrix}
                               1 \\
                               \mathbf{t} \\
                               \mathbf{t}^{2} \\
                               \mathbf{t}^{3}
                             \end{bmatrix} 
                        [1\, \mathbf{y}_{bkg}\, \mathbf{y}_{bkg}^{2}\, \mathbf{y}_{bkg}^{3}] \right)
\end{equation}

where $vec()$ denotes the vectorization operation, which unrolls a matrix into a vector, and $\mathbf{X}_{bkg}$ is a 2D vector with shape $(l \times n_{bkg}, 16)$. 
We find the best fitting model using linear least squares \citep[similarly as shown in][Appendix B]{2021AJ....162..107H}.
The resulting background model for each pixel and time $\mathbf{\hat{f}}_{bkg}$ is given by:

\begin{equation}\label{eq:bkg_model}
    \mathbf{\hat{f}}_{bkg} = \mathbf{X_{bkg}}\cdot \mathbf{\hat{w}}
\end{equation}

where $\mathbf{\hat{w}}$ are the best-fitting weights and $\mathbf{\hat{f}}_{bkg}$ denotes the best-fitting flux time series for the background pixels. 

The same weights can now be applied to a design matrix $\mathbf{X}$ of the pixel row positions of all pixels

\begin{equation}\label{eq:pix_dm}
        \mathbf{X} = vec\left(\begin{bmatrix}
                               1 \\
                               \mathbf{t} \\
                               \mathbf{t}^{2} \\
                               \mathbf{t}^{3}
                             \end{bmatrix} 
                        [1\, \mathbf{y}\, \mathbf{y}^{2}\, \mathbf{y}^{3}] \right)
\end{equation}

to evaluate the model at every pixel as $\mathbf{\hat{f}} = \mathbf{X}\cdot \mathbf{\hat{w}}$, where $\mathbf{\hat{f}}$ is the background flux time series of every pixel in the batch of TPFs.
Cadences where there is a significant, single deviation from this smooth model are identified and iteratively removed from the fit. 

Figure \ref{fig:bkg_model} shows the column-wise binned flux data as a function of pixel row and time for both the data (left) and model (center), and average flux (right). 
The model is able to capture the rolling band signal at the end of the quarter moving vertically in the CCD (vertical pale blue lines). 

To enable reproducibility, we package this model as a standalone simple Python package \emph{kbackground}\footnote{\url{https://github.com/SSDataLab/kbackground}}. 

\begin{figure*}[htb!]
    \centering
    \epsscale{1.15}
    \plotone{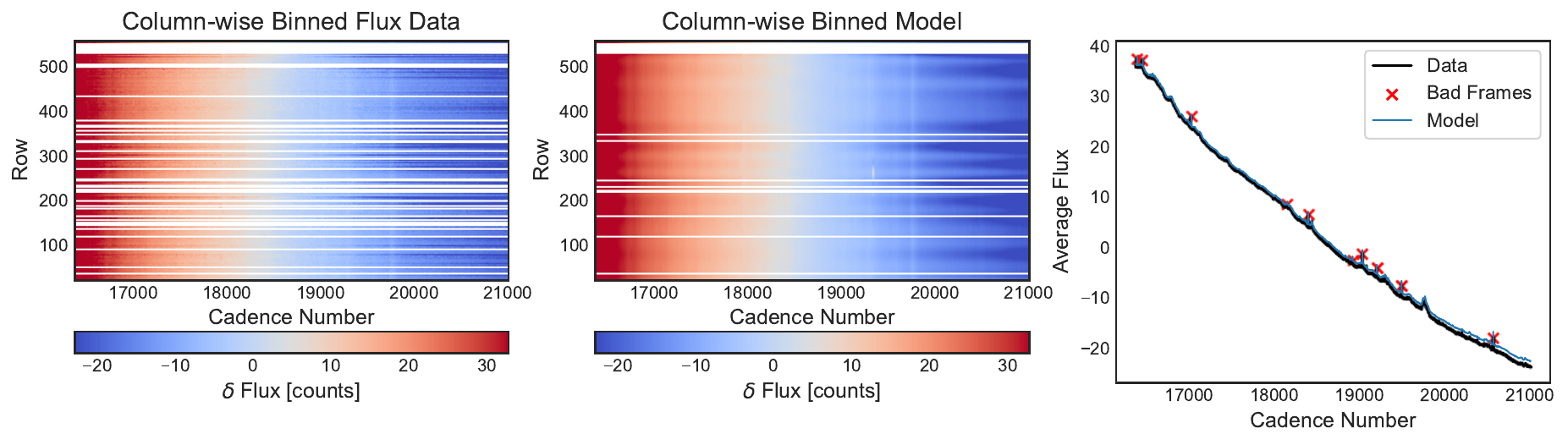}
    \caption{Example of background model estimated using the \texttt{kbackground} package for a stack of 200 TPFs in channel 44 observed during quarter 5. The left (center) panel shows the pixel flux data (model) as a function of time, each row on the y-axis represents the median average time-series across column pixels. The color maps the flux deviation from the mean value across all cadences. Because the rolling band signal is more prominent in the row direction, we model it as a function of time and pixel row, and average over the column dimension. 
    The right panel shows the flux value averaged from all background pixels as a function of time, the data is shown in black and the model in blue, red markers are cadences with an extra offset in the model due to outlier.}
    \label{fig:bkg_model}
\end{figure*}

The background model is subtracted from the original flux data to model the rolling band and any background trend (e.g. see the slope in Figure ~\ref{fig:bkg_model})  for each pixel. In this way, we obtain a zero-centered flux time series.

This rolling band model is adequate for our purposes but could be improved; we use only pixels that have no significant flux and do not model the sources simultaneously with the background. 
We use a simple spline model with fixed knot spacings rather than, for example, a Gaussian Process approach where hyper-parameters could be estimated. 
Finally, we average over the column dimension, removing any possibility of modeling the rolling band in the orthogonal direction. 
If there is any residual trend in this dimension, we will average over it.

\subsection{Point Spread Function Model} \label{subsec:PSF_model}

We used the PSF models computed in \citet{2022AJ....163...93M} which used Kepler's FFIs to generate robust and detailed PSF models for each CCD channel and quarter. 
The FFIs are single-cadence (30-minute exposure) images over every pixel taken at the beginning, middle, and end of a quarter (approximately one per month). 
\citet{2022AJ....163...93M} computed PSF models using Gaia EDR3 \citep{2021A&A...649A...1G} sources with a limiting magnitude $G=20$ which lead to the use of $\sim 12\,000$ sources and $\sim 100\,000$ pixel data per CCD to fit the models. 
As noted in \citet{2022AJ....163...93M}, the PSF models were computed for quarters and channels where Kepler extended background (EXBA) masks are available, i.e. quarters 4 to 17 and all CCD channels but 5 to 8, therefore we computed missing models for all channels in quarter 0 to 3. 
The models are stored in a Zenodo \footnote{\url{https://doi.org/10.5281/zenodo.5504503}} repository and are fully integrated into the \texttt{psfmachine} API.

We evaluate the FFI PSF models in a pixel grid 10 times finer than the original Kepler pixel size of $4 \arcsec / pixel$ (i.e. $0.4 \arcsec / pixel$) to find the factor by which the model needs to be scaled such that it integrates to one on the pixel grid from the stack of TPFs. 
This scaling factor encodes a combination of two effects, the finite integration due to the instrument pixel scale and the differences between Kepler $K_{p}$ and Gaia $G$ filters, as the model uses Gaia $G$ band fluxes as prior \citep{2021AJ....162..107H}.

Figure \ref{fig:psf_model} illustrates the PSF model and its residuals for quarter 5 channel 37. This corresponds to the PSF model fitted with the respective FFI data and evaluated at the positions of 250 TPFs. 
The PSF is fairly round for channels in the center of the Kepler field (like channel 37) with a slight elongation in one axis.
The centroid of the PSF data is under $\sim0.3\arcsec$ in each axis, see red marker in Figure \ref{fig:psf_model} top left panel. 
The scene centroid offset is computed as the mean of the offset values in each cadence. 
These cadence centroid offsets are estimated by averaging each data point (pixel) distance to its source coordinates (Gaia R.A. and Decl.) weighted by the Poisson uncertainty estimate. 
See Figure 2 in \citet{2022AJ....163...93M} for a display of PSF models for all channels (quarter 5). 
This shows that in channels near the border of the field, the PSFs are significantly distorted, with elongation and characteristic spike patterns. 

\begin{figure*}[htb!]
    \epsscale{1.15}
    \plotone{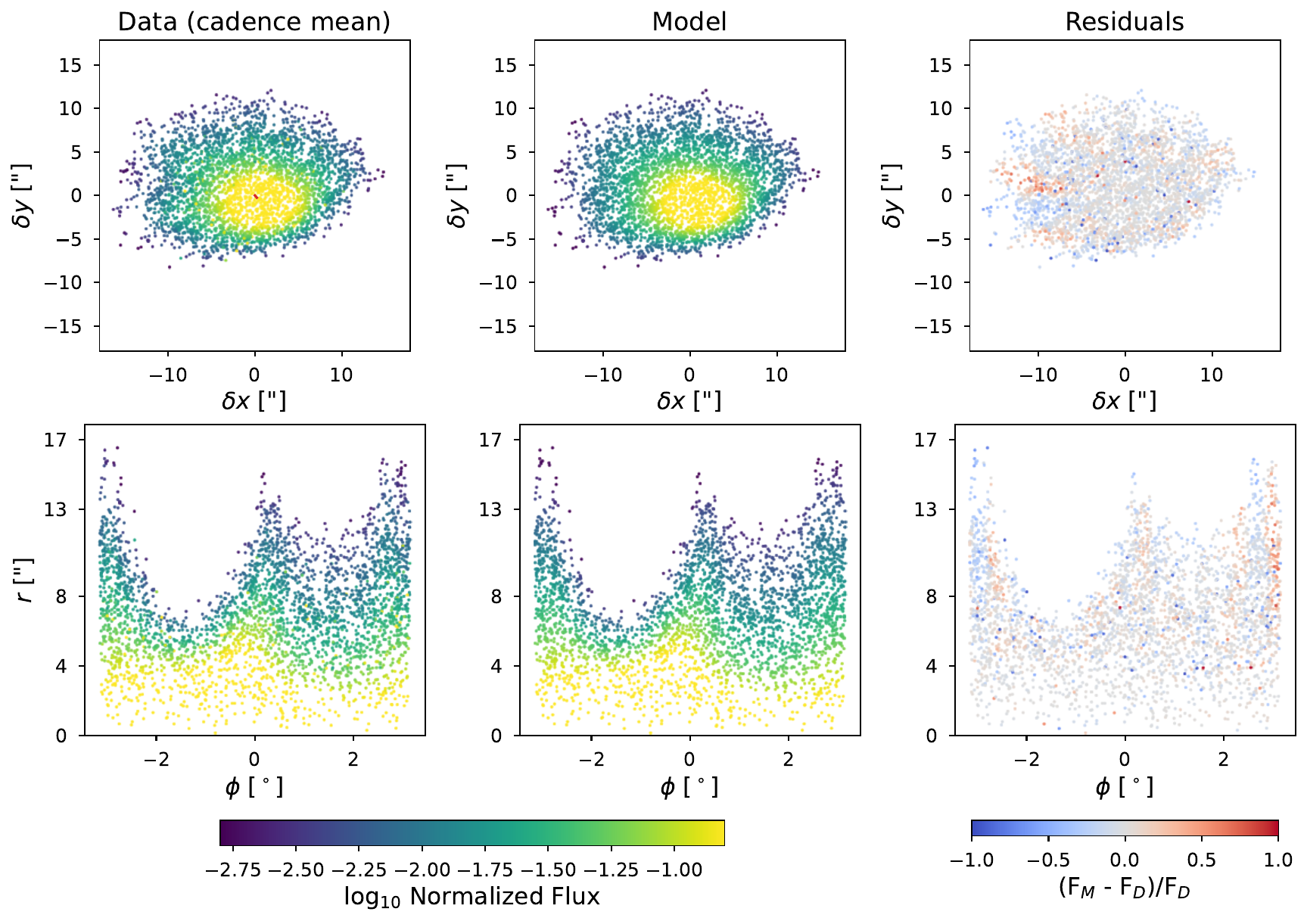}
    \caption{PSF data (left column), model (middle column), and residuals (right column) evaluated on a stack of 250 TPFs in channel 37 observed during quarter 5. The top row shows the data and model in Cartesian coordinates while the bottom row is in polar coordinates. Each data point corresponds to a pixel located $\delta x$ (radius) and $\delta y$ (angle) from the center of its corresponding source, and the color maps the normalized flux value. The red marker in the top left panel represents the magnitude of the PSF centroid offset which is $<0.8\arcsec$ with respect to the origin. A small fraction of data points have larger residuals due to remaining contaminated pixels that are clipped during model fitting.}
    \label{fig:psf_model}
\end{figure*}

\subsection{Correcting the Scene Motion} \label{subsec:per_model}

In the original LFD method, the differential velocity aberration effect \citep{2016ksci.rept....1V} is corrected using a spatial model and a third-order polynomial in time to create a time-dependent model \citep[see Section 4.5 in][for more details]{2021AJ....162..107H}. 
This time-dependent model is used to ``perturb'' the PSF model at each cadence, shifting the scene in its entirety. 
In this work we refer to the model that extracts flux time-series using the average PSF as the ``mean'' model, and the model that extracts flux time-series using the average PSF having been perturbed as the ``perturbed'' or ``corrected'' model.
The perturbed model accounts for small motions and slight changes in shape.
However, this method is only applicable if the PSF is fairly Stable and does not vary significantly. 
This is true for Kepler's primary mission observations, but not for K2 observations where the reaction wheel failure caused a systematic jitter motion in the spacecraft.

An alternative approach to correct the scene motion is to fit the PSF model to each frame separately, building a unique PSF model for every cadence. 
This process will mean fitting every variable in the PSF model ($<200$) for all of the $\sim4,500$ of \textit{Kepler} frames in a quarter. 
This adds up to $\sim900,000$ parameters.
With the number of usable pixel data ($\sim3,000$) in a stack of 250 TPFs this problem is not sufficiently constrained, resulting in a noisier estimation of the PSF overall. 
This problem could be overcome with more sources and more pixels, and fitting PSFs individually per frame becomes more tractable and beneficial but less computationally efficient. 
With the perturbation method approach, we fit a relatively small number of variables, $<200$ for the mean PSF model and $<1,000$ for the full perturbation model, leading to a well constrained and robust model.

We found that the third-order polynomial in time used originally in the LFD method can be too flexible and can introduce large-scale, spurious trends in the corrected light curves. 
This polynomial also does not address systematics other than large-scale motion, for example, the characteristic ``focus change'' signal that happens after the spacecraft downlink data. 

We implement an improved method to correct the scene motion and other instrumental signals. 
To find a reasonable solution we explored several approaches and their combinations. 
i) The centroid positions in each axis as basis vectors. 
These were either the mission-defined positional correction vectors \citep[POS CORR, ][]{2016ksci.rept....2V} or the centroid shifts computed by \texttt{psfmachine} via momentum method (average weighted by the Poisson noise).
ii) The components of common trends between the source pixels. These were estimated using principal component analysis (PCA) of the set of pixels belonging to aperture-extracted pixel time series of sources. 
iii) The mission Cotrending Basis Vectors \citep[CBVs, ][]{2016ksci.rept....2V}.
The CBVs were built by the mission pipeline \citep{2012PASP..124.1000S}. 
CBVs are built from the common trends across sources and contain multiple instrument systematic trends in sixteen basis vectors, systematic such as centroid shift due to reaction wheels adjustment, focus change due to data downlink, and others. 
For Kepler, single-scale CBVs are available in MAST archive and combine different time-scales systematics, e.g. long time scales to capture trends such as differential velocity aberration or short-term to capture focus change. Based on our investigation, we find of the three approaches CBVs perform the best to remove velocity aberration and focus change without introducing spurious signals. We find using the first four CBV vectors is sufficient to address the instrumental signal while keeping the dimensionality of the matrices low, and therefore computationally efficient. 
We apply a 2-day window smoothing b-spline function to each CBV vector to avoid introducing high-frequency noise from the CBVs into the corrected light curves. 
This smoothing step accounts for data discontinuity such as time gaps and value jumps.
Figure \ref{fig:cbvs} shows an example of the first four CBV components for channel 37, quarter 5.

\begin{figure}[ht!]
    \centering
    \epsscale{1.15}
    \plotone{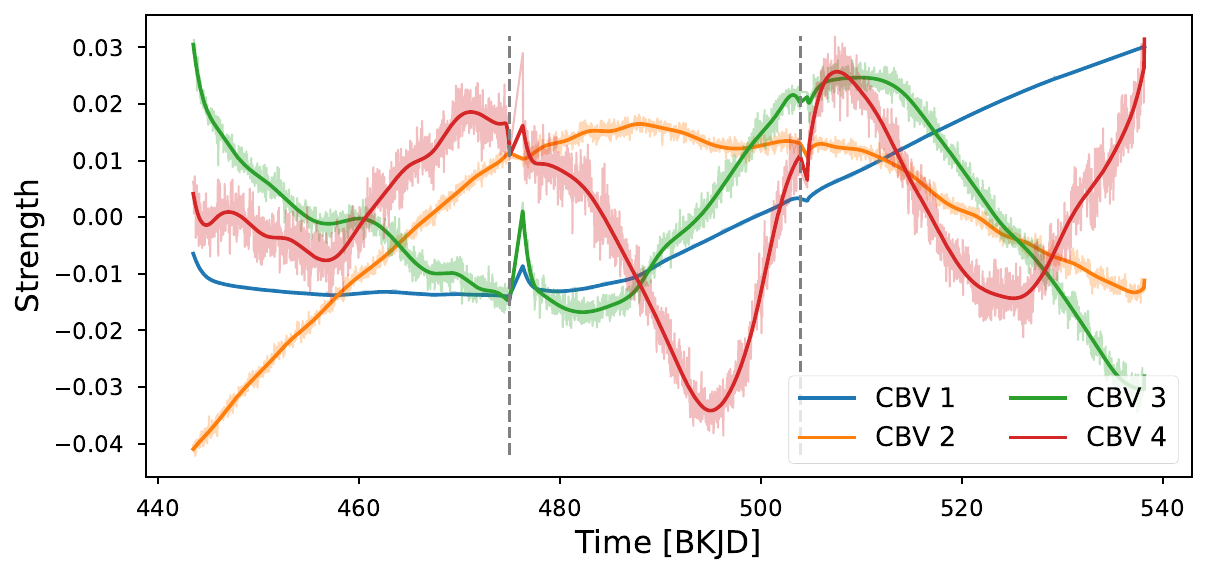}
    \caption{Example of first 4 CBV vectors for channel 37 and quarter 5. The colors are the different components, light colors are the original vector values, bold lines are the smooth version using a b-spline function, and vertical dashed lines show data gaps after the spacecraft performs data downlink.}
    \label{fig:cbvs}
\end{figure}

While our work and the \textit{Kepler} pipeline both use CBVs to address long-term trends, the methods used by each work are different.  
The \textit{Kepler} pipeline used CBVs to detrend each source individually \citep{2016ksci.rept....2V}. 
In our work, we apply CBVs to find the correction to the PSF model to best fit all sources in the batch of TPFs simultaneously, which is frequently $>400$ sources. 
Fitting multiple sources prevents overfitting the velocity aberration for an individual source. 
Additionally, in our method, we fit a low-resolution model in time to improve computational efficiency.
We use the CBVs to build our ``perturbation matrix'' which is then applied to the mean PSF model in all frames to track its changes.  
Figure \ref{fig:perturbed_model_space} shows an example of the perturbation matrix. This matrix is multiplied into the PSF model in order to change it to best fit the data in time.
The PSF changes from wider with significant wings to a narrower PSF, which the perturbation matrix is able to capture, see \cite{2021AJ....162..107H} for more discussion.

To build our perturbation matrix we bin data in time. 
This binning keeps the matrix small, thereby making memory usage and computing time low. 
By binning we reduced the time resolution from $\sim 4,500$ frames in time to 90 frames. 
Once the binned version has been fit to the data to find the best fitting weights, the model can be evaluated at all the cadences. 
By binning the data in this way we are assuming the motion is smooth and uniform, and that any differences between the data and this model are Gaussian distributed, which holds largely true during Kepler's primary mission observations. 
In contrast, data from the K2 mission exhibit a strong pattern due to the roll motion of the spacecraft, therefore this binning approach may not be adequate.

The binned time sequence for each data point of the perturbed model is shown in Figure \ref{fig:perturbed_model_time}. The figure shows the changes in time of each uncontaminated pixel used to fit the model for a batch of 250 TPFs on channel 37 during quarter 5. The data in this figure are mean normalized, causing there to be a turnover point close to index 50. 

The common trend (red to blue) is due largely to velocity aberration, focus change is evident as vertical clear stripes. Note that the magnitude and ``sign'' of this trend are different for each pixel, depending on whether a pixel is close to the center of a source or the wings, and whether the pixel is on the leading or lagging side of the target as it moves due to velocity aberration. The magnitude of the effect is commonly $\approx$20\%. 

Our time series model is shown in the middle panel, and is built from the PSF model for each source which has been perturbed and then fit to the image data to find the source flux.  After the removal of the perturbed model in our approach, the pixel time series residuals markedly improved to $\approx 2 \%$.

Our perturbation matrix results in a well-regularized model that preserves real physical variability, such as stellar activity or long-period variables, and removes most of the systematic trends due to velocity aberration and focus change. This is crucially different to the \textit{Kepler} Pipeline approach, as we are using the pixel data to inform our fit of the systematics and prevent overfitting.
See Appendix \ref{appx:lc_ex_lpv} for a direct comparison of long-period variable (LPV) light curves extracted with Kepler's \texttt{PDCSAP} and our PSF photometry.

\begin{figure}[htb!]
    \centering
    \epsscale{1.15}
    \plotone{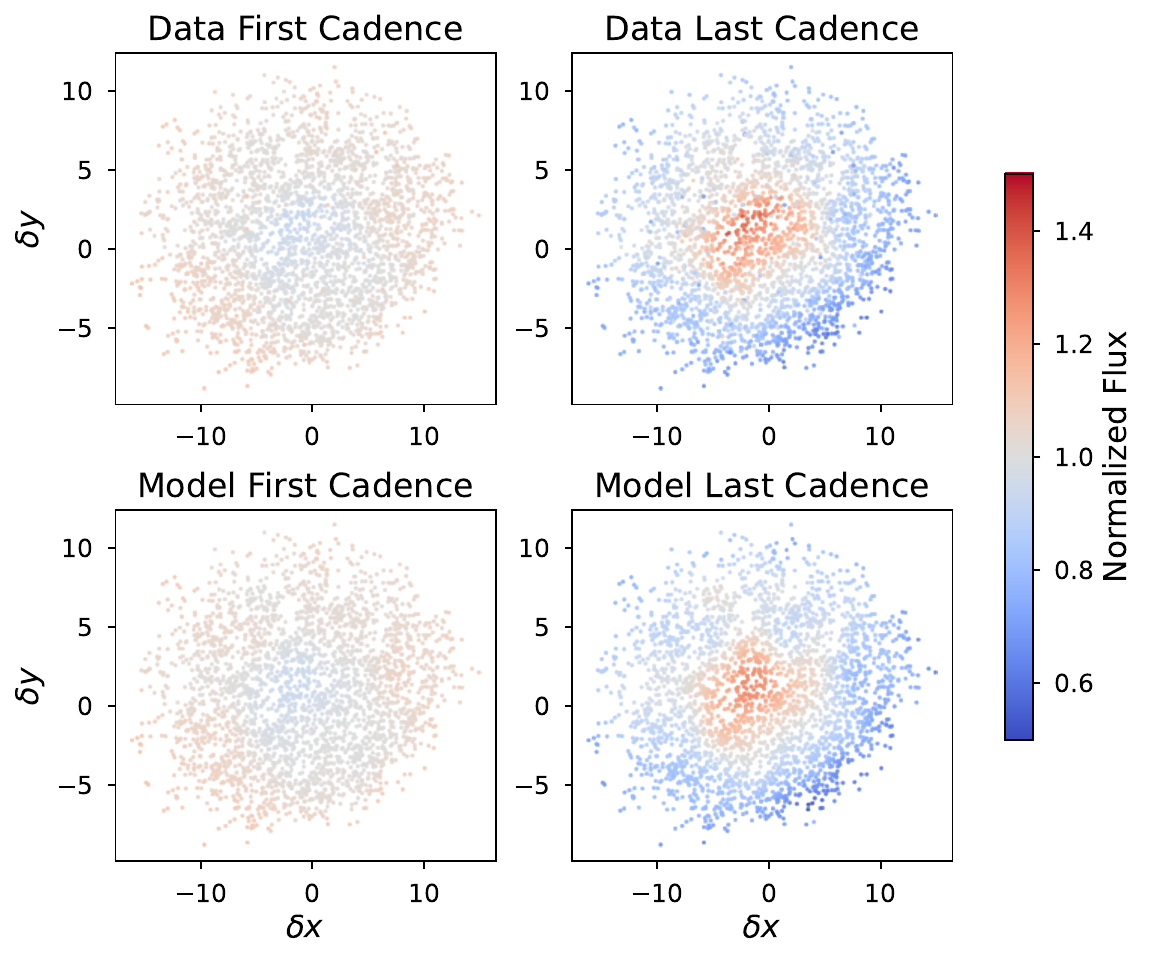}
    \caption{
    Perturbed PSF model for a stack of 250 TPFs in channel 37 quarter 5. The 4 panels show the spatial distribution of data points (top row) used to fit the perturbed model (bottom row) at the first (left column) and last cadence (right column). The shift in color shows the PSF profile change due to velocity aberration and other effects within the observing quarter. }
    \label{fig:perturbed_model_space}
\end{figure}

\begin{figure*}[htb!]
    \centering
    \epsscale{1.15}
    \plotone{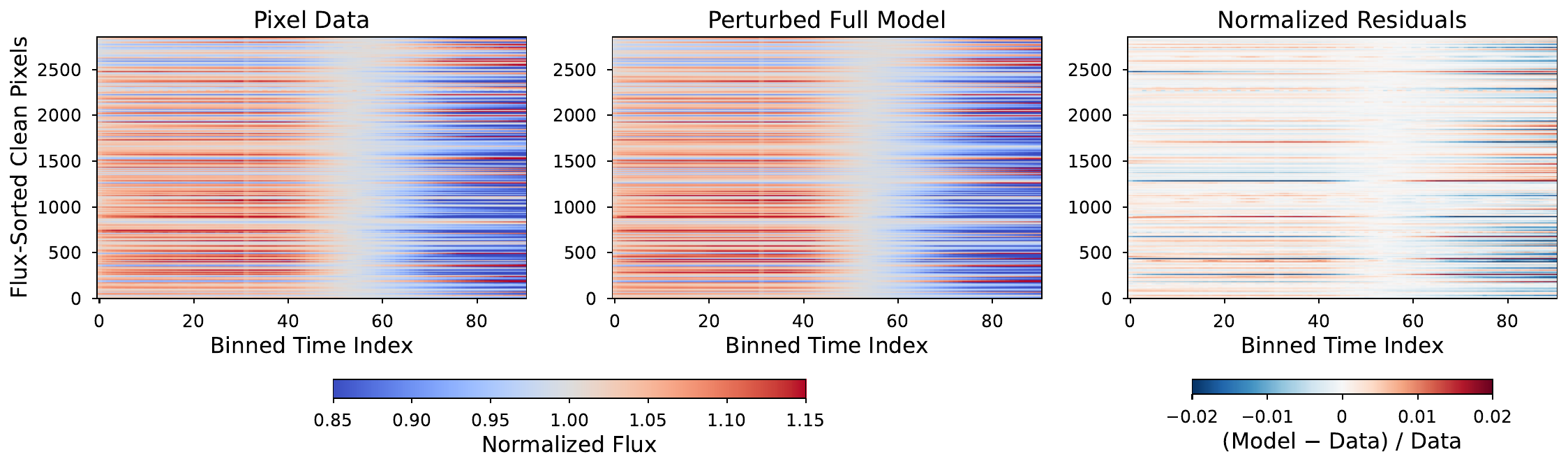}
    \caption{
    Mean normalized pixel time-series for uncontaminated pixel data in 250 TPFs in channel 37 quarter 5 (same as Figure \ref{fig:perturbed_model_space}). The 3 panels show the time series of each pixel data (left) after binning (as described in \ref{subsec:per_model}), the perturbed full model based on the \textit{Kepler} pipeline CBVs (middle), and the normalized residuals (left). 
    The pixel time series are sorted by flux value with fainter pixels at the bottom. 
The middle panel shows the "perturbed" version of the PSF model fit for all sources. Our model results in flux time series for each pixel, where the sources in the pixels are fit for flux at each time.
    The perturbed version of the full model is a good estimate of the data, as shown by the small residuals ($\le 2\%$). Data points with significant residuals are either remaining contaminated pixels or belong to true variable sources.
    }
    \label{fig:perturbed_model_time}
\end{figure*}


\subsection{Flux Priors and Iteration} \label{subsec:priors}

Since the LFD method uses linear modeling to fit the flux data, the solver can yield negative solutions that are mathematically correct. 
As negative flux values for stars are non-physical, we use narrowing priors to ensure target flux remains positive. 
As discussed in \cite{2021AJ....162..107H}, to estimate the flux of a source we solve the linear equation $\mathbf{\hat{f}} = \mathbf{S} \cdot \mathbf{v}$, where $\mathbf{\hat{f}}$ is our estimate of the pixel flux data, $\mathbf{S}$ is the PSF model, and $\mathbf{v}$ is a vector representing the intrinsic flux value of a source. 
Each source has a prior which is defined as a Gaussian with a mean ($\mathbf{\mu_v}$) and a standard deviation ($\mathbf{\sigma_v}$). 
$\mathbf{\mu_v}$ and $\mathbf{\sigma_v}$ are set as the Gaia G-band flux ($F_{G}$) and $10 \sqrt{F_{G}}$, respectively. 
The latter is a Poisson noise estimate, with a fairly wide prior.
In cases where $\mathbf{\hat{f}}$ for a source is negative, we narrow the priors for that source and its neighbors (up $5 \arcsec$ apart) by reducing $\mathbf{\sigma_v}$ by a factor of 2, constraining the fit. 
This narrowing is repeated three times, and any remaining negative sources are dropped from the source catalog.
Then a final fit is done with only the remaining positive sources.
While narrowing priors could potentially dampen intrinsic source variability for extreme cases, we find this approach to be adequate. 
With this iteration process, we are able to reduce the number of sources that return negative flux to about $2-5\%$ depending on how crowded is the area. 
Ultimately, $<5\%$ of the input sources are removed from the catalog due to negative flux estimations.


\section{Results} \label{sec:results}

This work presents a light curve catalog with $606,900$ sources. Our light curve files provide three main types of photometry; "aperture photometry", "mean PSF", and "corrected PSF"; as well centroid estimations, background model, and chi-square time-series from the PSF model. 
Both mean and corrected PSF are computed with the methods explained above. 

\begin{itemize}

\item  ``aperture'' photometry is computed from an aperture mask estimated as in \cite{2022AJ....163...93M}, which optimizes contamination and completeness of the flux within the mask.

\item  ``mean PSF'' photometry uses only the shape model loaded from the corresponding FFI and evaluated in the TPF stack data.

\item  ``corrected PSF'' photometry is obtained from the perturbed model fitted with the observed cadences in the TPF stack.

\item centroid vectors are computed by correcting the Gaia coordinates with offsets estimated using the momentum method (weighted average by Poisson uncertainty estimate) at every cadence. 

\item  background flux corresponds to the sum within the aperture of the model described in Section \ref{subsec:bkg_model}

\item chi-square time-series corresponds to $\mathbf{\chi}^2 = \sum{(\mathbf{f}_{model} - \mathbf{f}_{data})^2/\mathbf{f}_{data}}$, where $\mathbf{f}_{model}$ is the perturbed PSF estimate of the pixel data, $\mathbf{f}_{data}$ is the pixel flux, and the sum is over the pixel corresponding to the source. Chi-square time-series can be used both to diagnose where our extracted time-series may be imperfect, and any instances where the model does not fit well due to a changing PSF shape \citep{2021AJ....161...95H}.

\end{itemize}

Our light curve files contain the per-quarter light curves with the aforementioned measurements. 
Additionally, we provide a stitched version that contains the aperture, corrected PSF, and a flattened version of the PSF photometry.
The latter was flattened with a 2-day window b-spline function, designed to better enable the community to perform transit searches.
Appendix \ref{appx:lcfs} provides a specification of the content of the light curve files.

\subsection{Light Curve quality} \label{subsec:quality}

To assess the quality of the photometric extraction and performance of the light curves presented in this work, we produce a series of metrics. 
This section details the extraction of quality metrics for both types of photometry (aperture and PSF) as well as noise metrics to measure the light curve accuracy.

\subsubsection{Quality Metrics} \label{subsubsec:extraction}

During the light curve extraction process, we compute two aperture quality metrics and three quality PSF metrics. 

Similarly to the \textit{Kepler} pipeline we compute FLFRCSAP and CROWDSAP, as described in  \citealt{2022AJ....163...93M}.
FLFRCSAP is the fraction of target flux contained in the photometric aperture over the total target flux. 
CROWDSAP is the ratio of target flux relative to the total flux within the photometric aperture including contaminating sources.
These two metrics are computed using the evaluated PSF model on every source.
It is important to highlight that extracted sources with only partial coverage in the pixel data, (i.e. sources partially outside of the pixel cutout) could have overestimated FLFRCSAP and CROWDSAP values. 
FLFRCSAP can be lower because it is estimated only with recorded pixel data, while the CROWDSAP values could not account for contaminants further than $4\arcsec$ away from the TPF edge. 

We generate three new metrics to describe the quality of our light curves: 
\begin{itemize}
    \item PSFFRAC: how much of the total expected PSF was saved in the TPF (values of 0 to 1). Sources fully enclosed in a TPF will have values near 1 (because finite integration values are slightly lower than 1). Background sources that are partially on the TPF have values between 0 and 1. 
    \item PERTRATI: the ratio between the average flux from the mean model, and the average flux from the perturbed model. Sources with stable perturbed model have values close to 1. Values significantly different than 1 suggest a poor perturbation model mostly due to sparse fit data for the source. 
    \item PERTSTD: the ratio between the standard deviation of the mean model, and the standard deviation of the perturbed model. Small values indicate a stable perturbed model that does not introduce large variations to the extracted light curve.
\end{itemize}

From the PSF model, we estimated the object PSF fraction (PSFFRAC) on the pixel data, this is how much of the expected PSF was saved in the TPF. 
By design, the Kepler targets have the entire PSF inside the TPF. 
Background sources can have partial PSF, especially objects near or outside the TPF edges. 
For these sources, the PSF fraction can also vary between observation seasons due to the change in the spacecraft pointing or changes in TPF size. 
This can yield to changes in photometric accuracy, leading to noisier light curves when the PSF fraction decreases. 
Due to this effect, we only provide stitched light curves from quarters with a PSF fraction larger than 0.5 to avoid the use of low-quality photometry. 
We still include all the extracted quarters in the light curve FITS file, regardless of low PSFFRAC values.

To measure the effects of introducing the perturbation PSF model, we compare it against the mean PSF model estimated early in the process. 
We computed the ratio between the mean PSF and the perturbed PSF and took the mean (PERTRATI) and the standard deviation (PERTSTD). 
These metrics measure how much the perturbed model deviates from the mean PSF and the introduced variance. 
Both metrics can be used to filter light curves where the perturbed model introduces artifacts. 
In this case, we recommend defaulting to the photometry fitted with the mean PSF model.

\subsubsection{Photometric Noise} \label{subsubsec:cdpp}

\begin{figure}[htb!]
    \centering
    \epsscale{1.15}
    \plotone{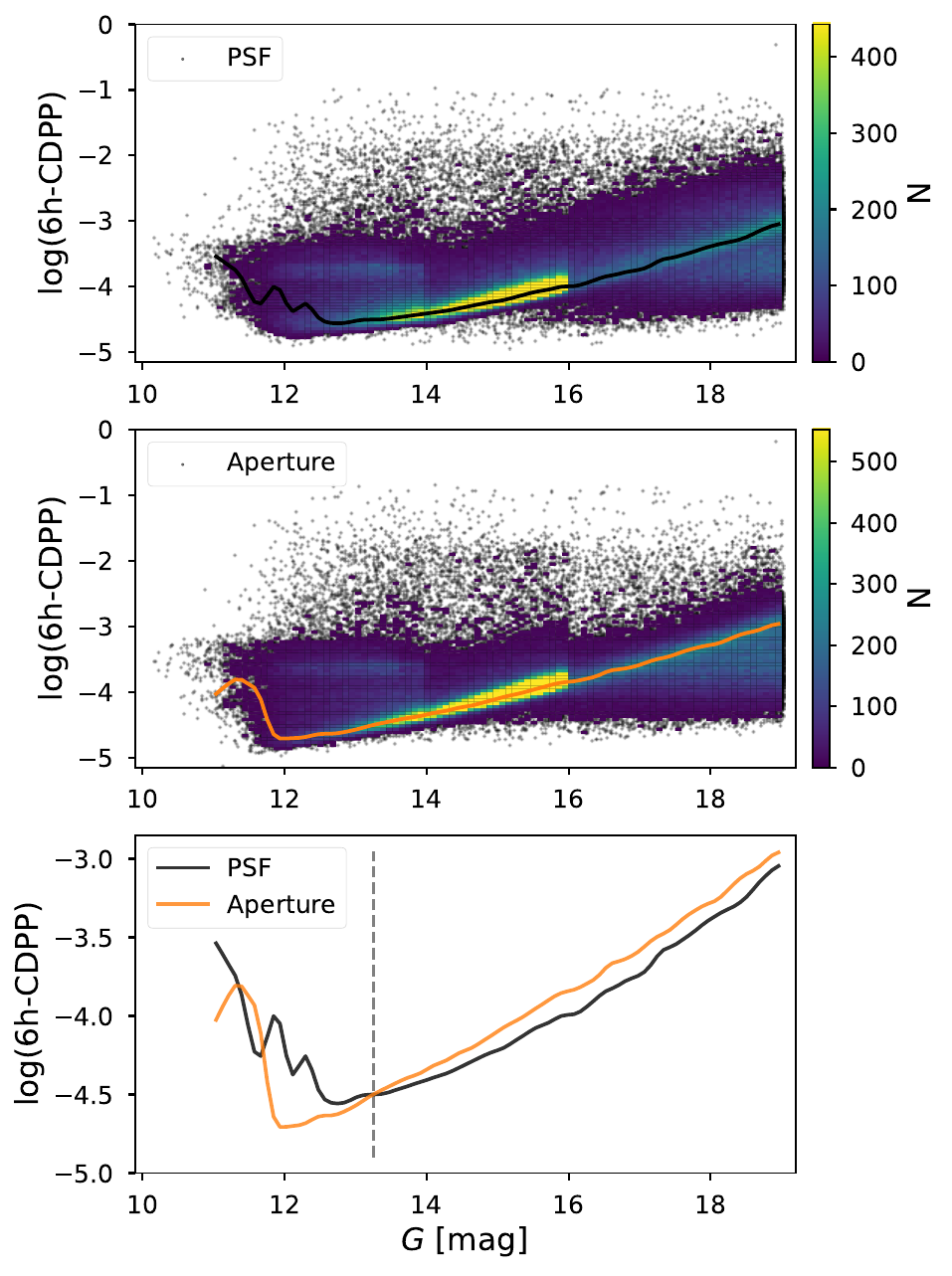}
    \caption{6-hour estimated CDPP in parts-per-million as a function of G-band magnitude from the perturbed PSF (top) and aperture (middle) light curves. CDPP estimates the noise properties of \textit{Kepler} data, where lower values indicate a more precise light curve. The bottom panel shows a direct comparison of PSF and aperture CDPP values. The black and orange lines follow the CDPP distribution maximum in a brightness bin (0.1 magnitude width). 
    As expected, the noise increases as flux decreases, reaching the 100 ppm value for objects fainter than $G = 16$, where the Kepler target selection bias is shown as a hard cutoff in bin counts.
    The gray dashed line shows the point at 13.25 mag where on average PSF photometry achieves lower CDDP values than aperture photometry.
    See Section \ref{subsubsec:cdpp} for details.}
    \label{fig:cdpp_v_mag}
\end{figure}

The \textit{Kepler} pipeline introduced a metric to estimate the noise quality of light curves, this is the Combined Differential Photometric Precision (CDPP). 
We use the implementation of the estimate of CDPP metric included in the \texttt{lightkurve} Python package, which implements a simpler version of CDPP \citep{2011ApJS..197....6G, 2016PASP..128g5002V}.
We compute this metric for every extracted source in this word. 
Figure \ref{fig:cdpp_v_mag} shows the estimated CDPP values as a function of G-band magnitude for all sources with a PSF fraction larger than $0.5$. 
A large number of sources, brighter than $G = 16$ have CDPP values under 100 ppm, which is comparable to values estimated from the PDCSAP light curves computed by the \textit{Kepler} pipeline. 

Figure \ref{fig:cdpp_v_mag} shows there is a turnover point where PSF photometry becomes more accurate than aperture photometry, at approximately G=13.25, indicating a significant benefit in precision. 
The high-density horizontal ridge at $log_{10}(\textrm{6h-CDPP}) \sim -3.5$ between 12th and 14th magnitude shows CDPP values about one order of magnitude larger than the main trend. 
An inspection of the Color-Magnitude diagram (CMD, see Section \ref{subsec:demographics} for details) showed that these sources correspond to red giant stars near the horizontal branch, see Figure \ref{fig:noisy_hb}.
This sample of red giants is about $1.6\%$ of the total catalog.

\begin{figure}[htb!]
    \centering
    \epsscale{1.15}
    \plotone{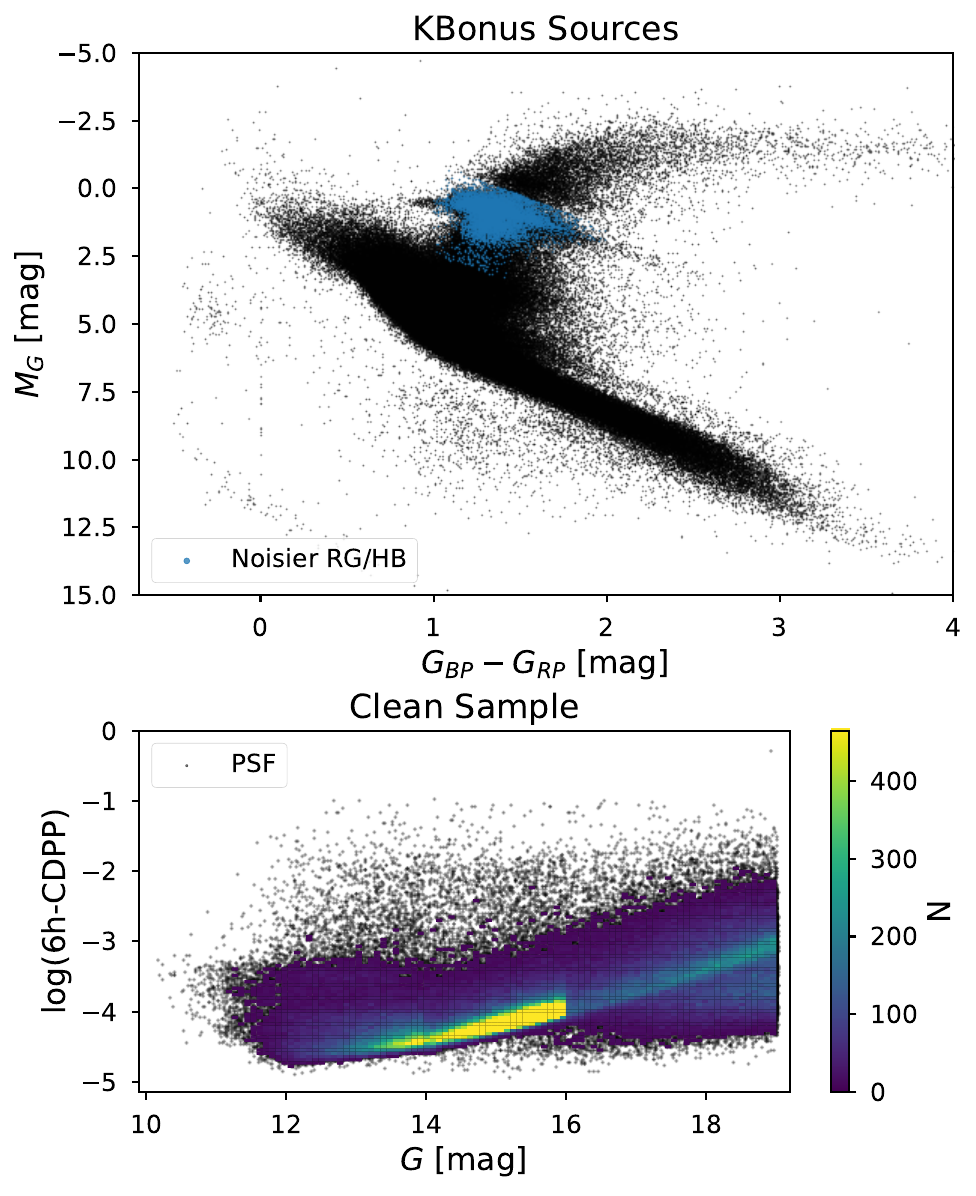}
    \caption{
    \textit{Top}: color-magnitude diagram of KBonus sources (black) and a sample of evolved red giant and horizontal branch stars (blue) that show consistently higher CDPP values with respect to the main distribution. 
    \textit{{Bottom}}: 6h-CDDP values as a function of G-band magnitude from the perturbed PSF model. The figure shows sources after removing the above-mentioned evolved stars. The distinctive high-density band around $log_{10}(\textrm{6h-CDPP}) \sim -3.5$ seen in Figure \ref{fig:cdpp_v_mag} is not present.
    }
    \label{fig:noisy_hb}
\end{figure}


\subsubsection{Light Curve Correlation}\label{subsubsec:corr}

Although the LFD photometry method presents many advantages such as computing speed, extraction of a large number of sources simultaneously, and ability to deblend contaminated sources, it suffers the problem of correlated light curves. 
Correlated time-series from different sources can occur when
\begin{itemize}
    \item Extremely close targets are difficult to separate, and the solution becomes close to degenerate
    \item The PSF model and/or source locations are incorrectly estimated
    \item The PSFs of each source vary in shape in a way that is not captured in the model, (e.g. sources of different colors have weakly different PSF shapes)
\end{itemize}

To assess when light curves are significantly correlated with each other, we compute the Pearson coefficient $r$ between pairs of light curves.

We found all pairs of time series within $60 \arcsec$ from each other and then removed the long-term trend from the light curves using a third-degree polynomial in time, (removing any long-term variability due to residual systematics, while preserving periodic variability).

We compute the Pearson correlation coefficient between the time-series pairs. 
Figure \ref{fig:corr_2d} shows the distribution of statistically significant (p-value $< 0.05$) coefficients as a function of pair distance. 
High values of $r$ indicate that the pairs are significantly correlated.
The majority of pairs have values $r < 0.15$, meaning no significant correlation. 
Pairs with values $r \sim 0.4$ demonstrated no visual correlation after inspection.
Almost all correlated pairs ($r > 0.5$) are within $25 \arcsec$, which relates to the typical size of a TPF ($\sim 5$ pixels across) meaning correlated pairs are likely found in the same TPF. 
Less than $1 \%$ of pairs fall in the correlated region ($r > 0.5$ and $d < 25 \arcsec$). 
Pairs with $r > 0.5$ beyond $25 \arcsec$ do not show correlated signals and the r values are likely due to remaining monotonic trends.
For a correlated pair, we assume the brighter source is the true variable, and the fainter gets contaminated. 
We opt to remove from our light curve catalog the faint source ($\sim 1 \%$ from the total data set) from every correlated pair.

\begin{figure}[htb!]
    \centering
    \epsscale{1.15}
    \plotone{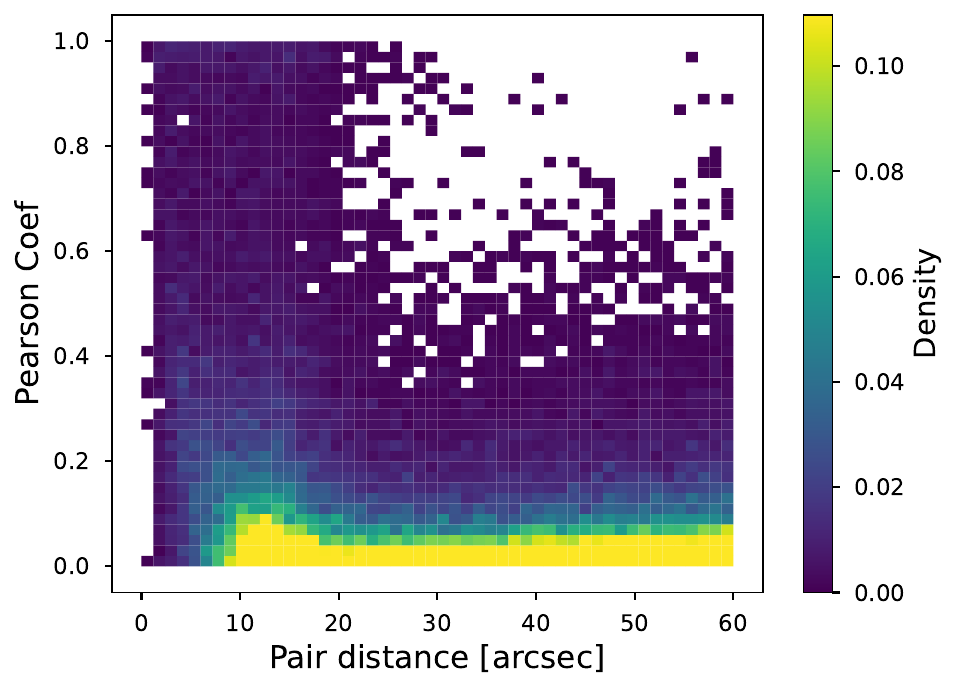}
    \caption{Pearson correlation coefficient between light curve pairs as a function of distance during quarter 5. Light curve pairs are made of every possible pair of sources within $60 \arcsec$. Higher coefficients mean a significant correlation between light curves. The break in density around $25 \arcsec$ is due to the typical size of a TPF. The high-density feature with $r < 0.2$ represents the majority of non-correlated pairs. The bulge feature between $7 - 20 \arcsec$ and $r < 0.3$ is due to the typical distance between pairs in a TPF, while the drop in density within $5 \arcsec$ is due to isolated sources in a TPF and the rejection of predicted negative fluxes (see Section \ref{subsec:priors}).}
    \label{fig:corr_2d}
\end{figure}

\subsection{Sources Demographic} \label{subsec:demographics}

This work presents the first catalog of light curves using observations from \textit{Kepler}'s prime mission nearly without a selection bias.
The number of new light curves ($> 400,000$) doubles the total delivered by the \textit{Kepler} pipeline ($\sim 200,000$). 
Figure \ref{fig:hr_diag} shows the color-magnitude diagrams (CMD) using Gaia DR3 photometric bands and distances computed by \cite{2021AJ....161..147B}. 
As a comparison, Figure \ref{fig:hr_diag}  shows \textit{Kepler} targets only (top left), new sources (top right), all sources in the catalog (bottom left), and the ratio between both samples (bottom right). 
The KBonus Background sample has significantly more sources around the main sequence region, particularly toward redder colors and the binary sequence. 
The number of new sources is smaller than \textit{Kepler}'s target for some evolved stars, particularly for luminous red giants, where the addition of new sources is $\sim10\%$. 
However, there is a significant increase in new low-luminosity red giants.
This reflects the target selection bias imposed by the Kepler mission that favored luminous red giants instead of cool, low-luminosity giants.

\begin{figure*}[htb!]
    \centering
    \epsscale{1.15}
    \plotone{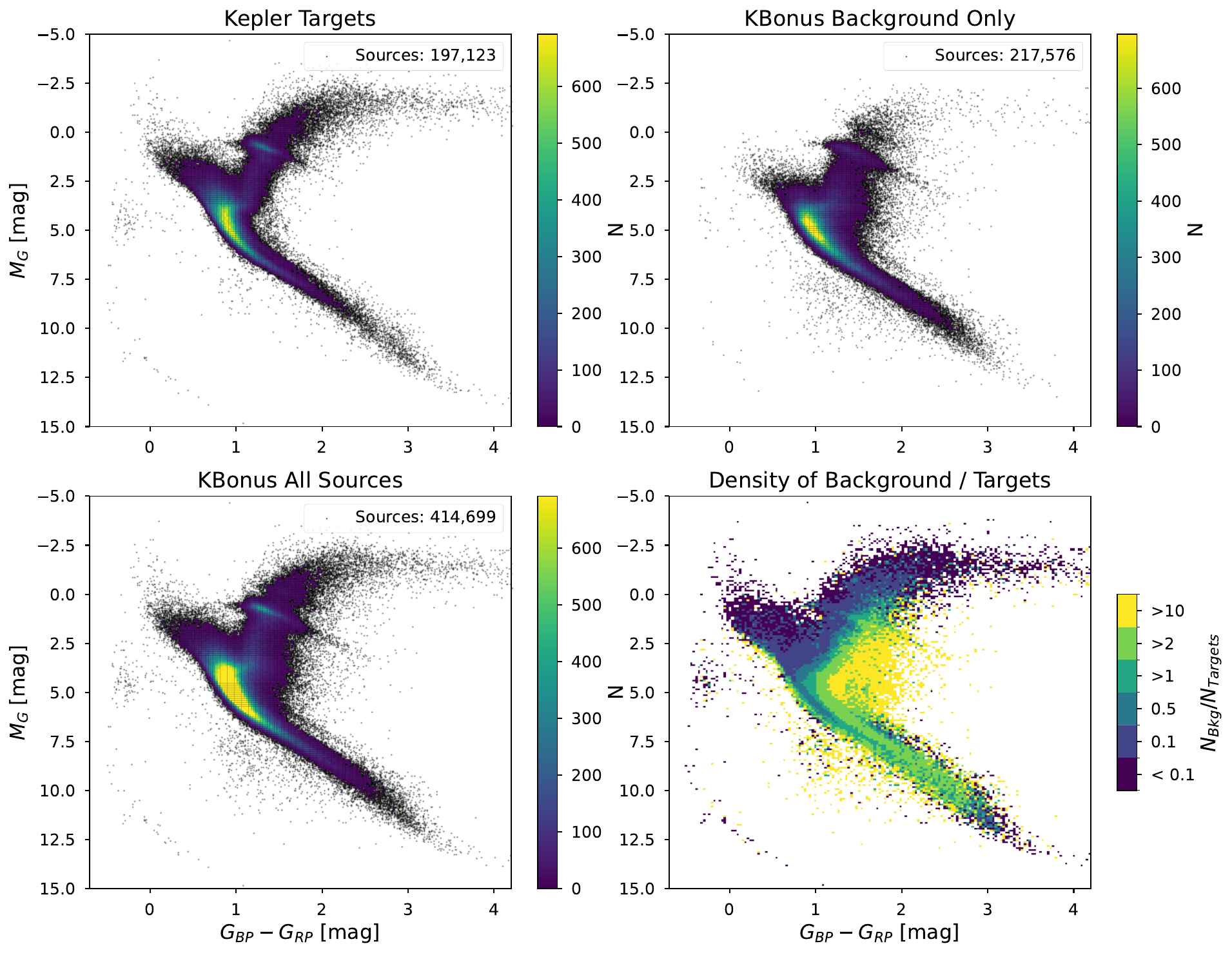}
    \caption{Color-magnitude diagrams of the KBonus catalog presented in this work. Only Kepler targets are in the top left panel, new background sources are in the top right panel, and the combination of both data sets is in the bottom left panel. 
    These three panels are 2D histograms with a 10 counts threshold and scatter data points elsewhere. 
    The fraction of new sources over Kepler targets in the CMD is shown in the bottom right panel. 
    The discrete colors map regions where the background sources are less than 10\%, or more than 10\%, 50\%, 100\%, 200\%, and 1000\% when compared to the number of Kepler targets.
    Only sources with Gaia parallaxes $> 0.001$ mas and valid BP and RP magnitudes are displayed. 
    Notable regions in the CMD are the red clump, the binary sequence, and the main sequence, where the addition of new sources is substantial. This is a result of an unbiased selection of sources in the KBonus catalog.
    }
    \label{fig:hr_diag}
\end{figure*}

The Kepler mission targeted approximately $3,700$ M-dwarf stars. 
In this work, we expand the catalog with almost $27,500$ new light curves in the M-type dwarf region of the CMD. 
We follow the prescription presented in \cite{2018arXiv180406982B} based on Gaia, WISE, and 2MASS bands (if available) to select potential M-dwarfs combined with cuts in the characteristic range of stellar temperature of m-type stars using Gaia's effective temperature.
Additionally, there are 50 new light curves in the white dwarf (WD) sequence, in addition to the previously extracted 41 WD Kepler targets.

\subsection{Confirmed Exoplanets} \label{subsec:exos}

We compared the estimated transit depth of previously confirmed Kepler exoplanets between Kepler's PDCSAP and our PSF light curves. 
We select all exoplanets with Archive Disposition `CONFIRMED' from the NASA Exoplanet Archive \citep{KOIs_cumulative}. 
To compare the transit depth measured directly on PDCSAP and PSF light curves, we performed a Box Least-square \citep[BLS,][]{2002A&A...391..369K} periodogram using a dense grid around the reported periods. The BLS method searches for periodic variability by fitting the data with an upside-down top-hat periodic model and it has been extensively used to analyze transiting signals \citep[e.g.][]{2011AJ....141...83P,2013PASP..125..989A,2015ApJ...806..215F}.
Both PDCSAP and PSF light curves were flattened beforehand to remove stellar variability using 2-day window b-spline functions while masking out cadences with transits.

\begin{figure}[htb!]
    \centering
    \epsscale{1.15}
    \plotone{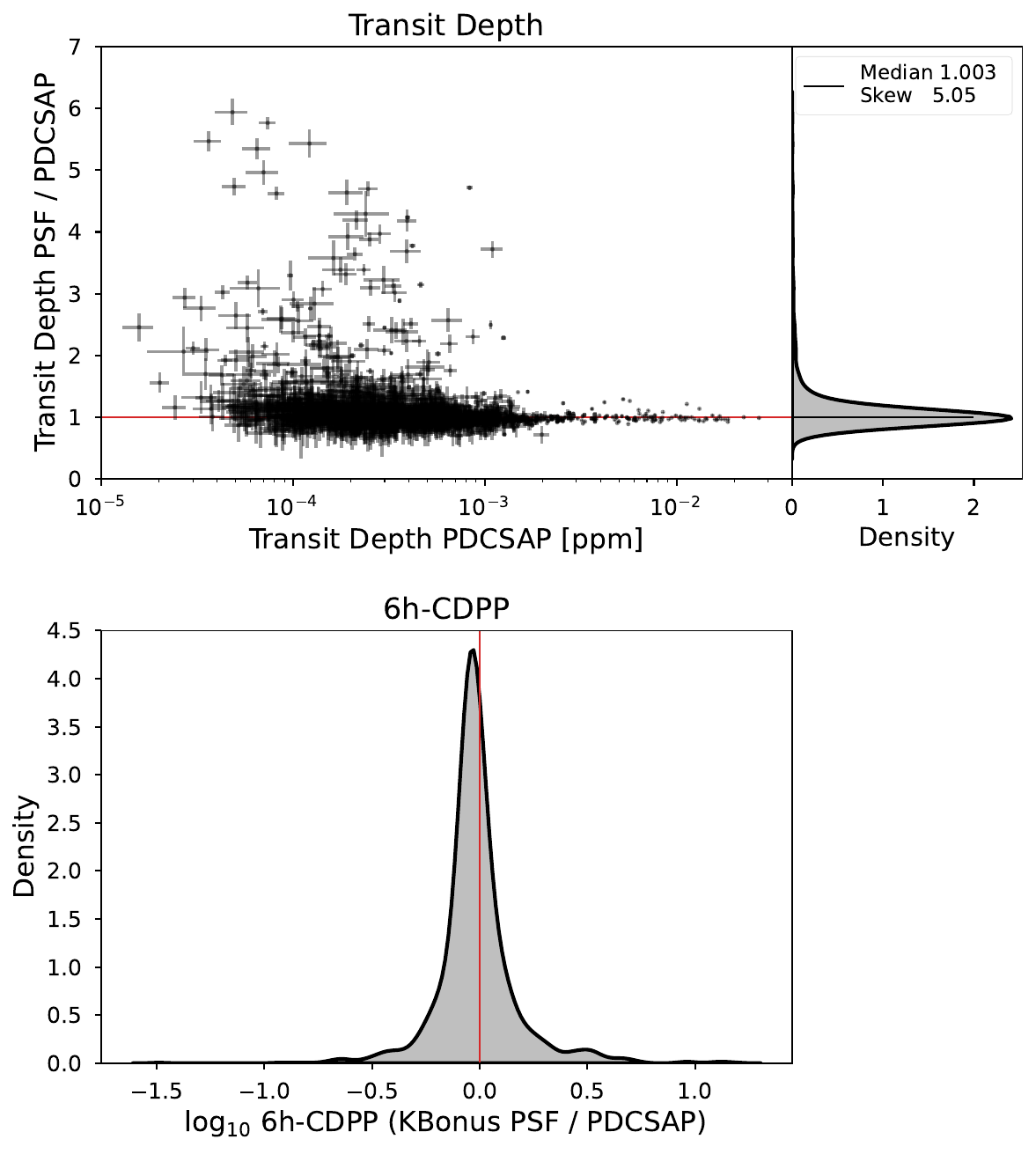}
    \caption{\textit{Top}: comparison of confirmed exoplanets transit depth measured from Kepler PDCSAP light curves (x-axis) as delivered on the NASA Exoplanet Archive, and the ratio (y-axis) of the transit depth measured in our PSF light curves and the PDCSAP light curves, from quarter 5 only. The right panel shows the distribution of ratios, which is centered near one with a median value of $1.003$ and slightly skewed ($5.05$) to values above 1. i.e. slightly larger transit depths are recovered by our PSF light curves, which is to be expected if some targets are contaminated.
    \textit{Bottom}: The density distribution shows the ratio in logarithm space of the CDPP values computed from the PSF and PDCSAP light curves. The red vertical line marks the ratio value 1 where both light curves have the same CDPPs. Negative values mean the PSF light curve has less noise, while positive values mean higher CDPP values than PDCSAP light curves. The distribution is offset towards negative values, indicating PSF light curves are less noisy.}
    \label{fig:transit_depth}
\end{figure}

Figure \ref{fig:transit_depth} shows the comparison in transit depth between both PDCSAP and PSF light curves. 
Overall the computed transit depths are consistent. 
A Skewness value of 5 for the ratio between depth values (PSF over PDCSAP) means the PSF light curves yield slightly deeper transits. 
This is expected, as some of the Kepler apertures could be contaminated by nearby sources, which is addressed by the PSF photometry. 
The 6h-CDPP values from both light curves are also consistent.

We measure the impact of the transit depth change on the estimated planet radius by scaling literature planet sizes by a factor which is the ratio between the radius estimation from the PSF and the PDCSAP light curves. 
We found a minor change in the exoplanet population towards a tighter distribution in planet size (see Figure \ref{fig:planet_size}), particularly for orbital periods longer than 20 days.
But no significant change in the population of global planet sizes. 
Further analysis of planet size populations will require complete exoplanet modeling using up-to-date host stellar parameters and Bayesian inference. 
We leave this analysis for a future study.

\begin{figure}[htb!]
    \centering
    \epsscale{1.15}
    \plotone{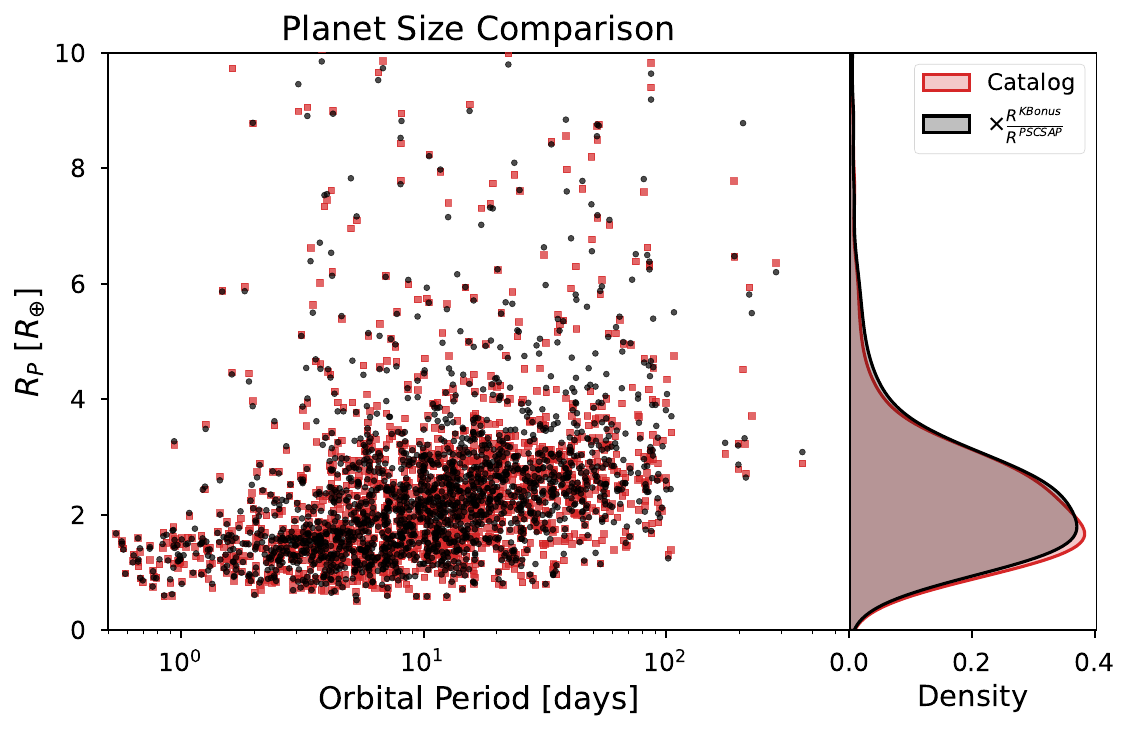}
    \caption{Planet radius as a function of the orbital period for the sample of confirmed exoplanets shown in Figure \ref{fig:transit_depth}. The red squares correspond to the planet radii reported in the NASA Exoplanet Archive. The black points are the planet radii scaled by the ratio between the measured radius from the PSF and PDCSAP light curves.
    The right panel shows the planet size density distributions in the same color scheme.
    A small change in the planet size distribution around $R_P \sim 2$ for periods longer than 20 days appears to tighten up the distribution.
    The significance of this effect needs to be tested with a complete exoplanet analysis, which is out of the scope of this work.
    The figure is intentionally limited to $R_P < 10$ in order to see the most populated region of the parameter space.
    }
    \label{fig:planet_size}
\end{figure}

\subsection{Revisiting False Positives KOIs} \label{subsec:kois}

To demonstrate the potential unlocked by this new light curve catalog, we examine a sub-sample of Kepler Object of Interest (KOIs) that were flagged as centroid offset false positive exoplanet candidates. 
These are light curves where it is likely the aperture is contaminated by a background eclipsing source, likely an eclipsing stellar binary, one of the main sources of contamination when searching for exoplanets via the transit method. 
Thanks to the use of a (nearly) complete source catalog (down to G magnitude 19th) and PSF photometry we are able to separate blended sources up to $1 \arcsec$. 
In this section, we present two KOI examples where we are able to separate the eclipsing sources. 
See Appendix Section \ref{appx:lc_ex_koi} for more KOI examples. 
A full analysis of all the false positive KOIs ($> 1800$) is left for future work.

\subsubsection{KOI 770.01} \label{subsec:koi_1}

KOI 770.01 is a false positive candidate with a reported transit period of 1.506 days and a $2211$ ppm depth, but it was flagged with centroid offset. 
The top panels in Figure \ref{fig:koi_1} show the Kepler PSCSAP light curve (red line) computed from a 4-pixel aperture mask (red mask on the pixel image). 
We generate two PSF time-series for this dataset, for the two sources blended in the image (shown in the pixel image by a black and blue marker).
Our PSF light curve (black line) of this source (black marker) does not show the transiting signal. 
The neighbor source \textit{Gaia DR3 2134870879540928896} (blue marker in the pixel image), is $1.36 \arcsec$ from the KOI and it is 2.6 magnitudes fainter.
The PSF photometry for the contaminant shows a clear eclipse signal at the same period (blue light curve). 
By the shape of the transit and its depth, the contaminant is a potential EB.
Our PSF photometry successfully separates both highly contaminated sources at high contrast.

\begin{figure*}[htb!]
    \centering
    \epsscale{1.15}
    \plotone{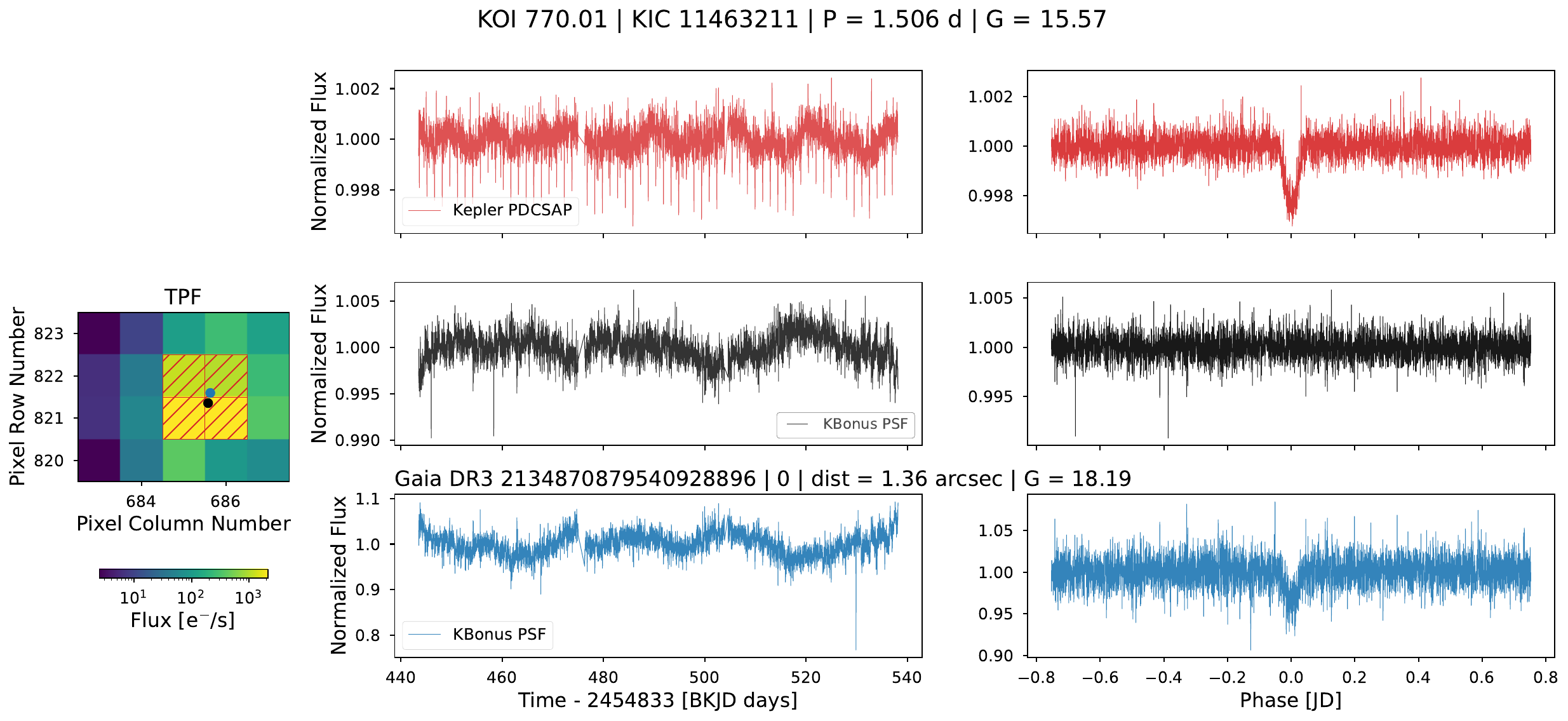}
    \caption{KOI 770.01 false positive due to centroid offset. The image shows an example cadence of the pixel data (TPF), highlighted in red is the aperture selected by the Kepler pipeline, the black dot shows the position of the candidate target, and the blue dot shows the position of the neighbor. The light curve panels show the time-based (middle) and period folded (right) time series for the Kepler PDCSAP light curve (top), our PSF photometry for the candidate (middle), and the contaminant (bottom) following the same color code as above. KIC or Gaia identifiers, period values, magnitudes, and pair distances are detailed in the panel titles.}
    \label{fig:koi_1}
\end{figure*}

\subsubsection{KOI 909.01} \label{subsec:koi_2}

Similar to the previous case, the KOI 909.01 is also a centroid offset false positive. 
Figure \ref{fig:koi_2} shows the pixel image and the light curve of the target and neighbor. 
The transit depth is $4147$ ppm with a period of 16.37 days. 
The source of contamination in the target aperture is a neighbor Kepler target, KIC 8256044, flagged as an EB. 
While these two sources are not highly blended, the separation between stars is $8 \arcsec$ (2 pixels), KIC 8256049 flux leaked into the candidate's aperture.
The PSF photometry is able to successfully deblend both light curves.
The difference in amplitude of the stellar variability seen between the Kepler PDCSAP and our PSF photometry for KIC 8256049 is mostly due to aperture contamination.

\begin{figure*}[htb!]
    \centering
    \epsscale{1.15}
    \plotone{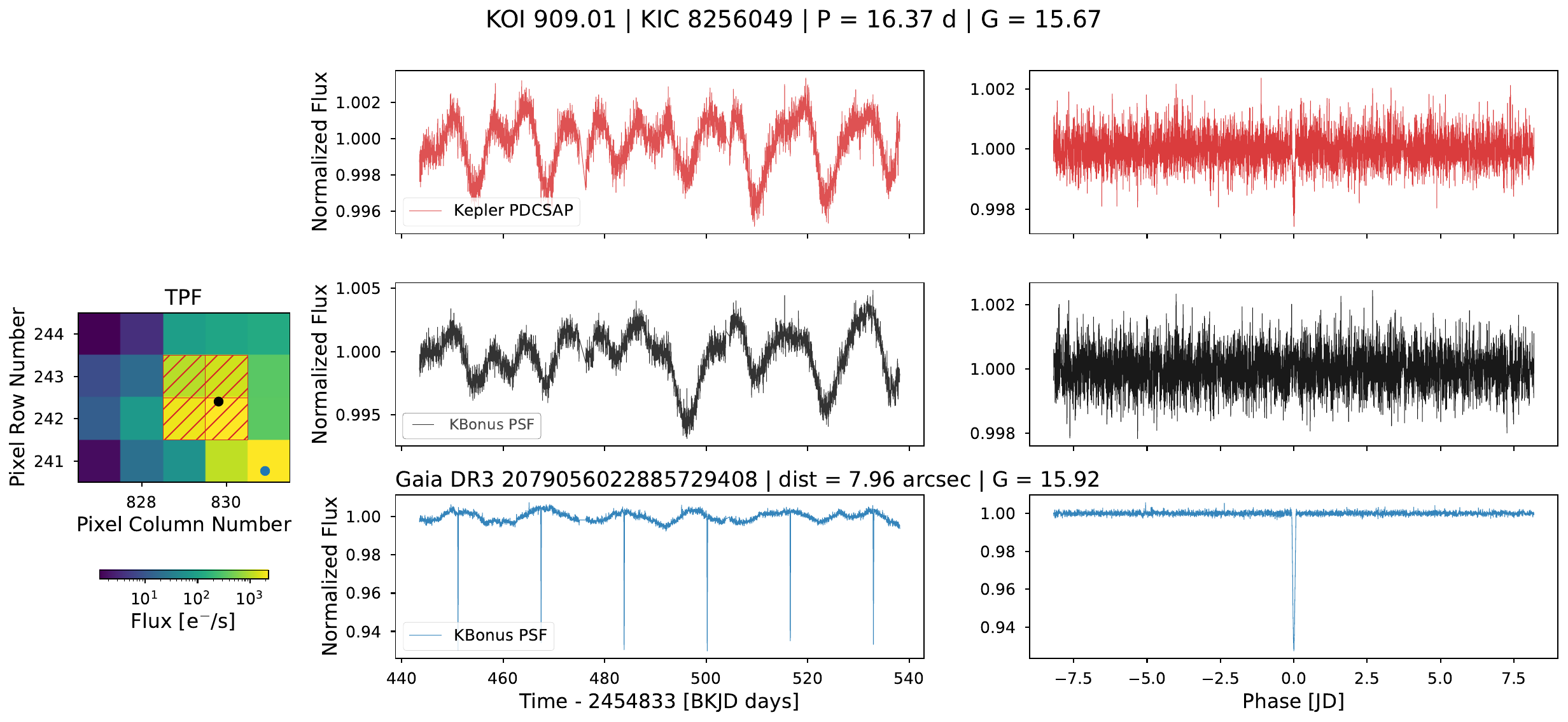}
    \caption{KOI 909.01 false positive due to centroid offset. Legends are the same as Figure \ref{fig:koi_1}. The contaminant is KIC 8256044 which was labeled as an EB.}
    \label{fig:koi_2}
\end{figure*}


\section{Limitations} \label{sec:limitations}

Due to the assumptions made throughout this work, some limitations for the PSF, the perturbation model, and the light curve arise. 
Here we list and discuss some of these limitations.

\begin{enumerate}

    \item The light curve catalog is limited to sources brighter than G band 19th magnitude and dimmer than magnitude 10th. 
    For blended sources within $1\arcsec$ only the brighter object was extracted while the fainter was removed. 
    Users can use the \texttt{psfmachine} API to extract sources outside the aforementioned ranges, although a fine-tuning of model parameters could be required for the PSF model to work outside the linear response range of the CCDs.
    
    \item Sources that are near the edge of the TPFs or outside of them are fitted using partial data. 
    Although PSF photometry is still able to extract them, the precision is not optimal. 
    Due to seasonal pointing accuracy, changes in the TPF shape, or high proper motion sources around the edge of the TPFs can have a different fraction of flux on the pixel data across quarters. 
    This is reflected as a change in photometry precision between quarters. 
    To minimize the risk of using subpar precision light curves we only stitched quarters with PSFFRAC $\geqslant 0.5$. 
    Users can still access the light curves for all quarters as they are provided in the multi-extension FITS files. 
    Additionally, aperture photometry, as well as the extraction metrics FLFRCSAP and CROWDSAP, for these partial sources are underestimated.
    
    \item The PSF models are fitted, as described in \cite{2021AJ....162..107H}, by solving a linear model as a function of positions. 
    This approach does not account for the change in shape due to source brightness. 
    Brighter sources can have a slightly different profile shape than fainter sources. 
    Although we evaluated the option of a flux-dependent PSF model, these changes were noticeable in the outer regions of the PSF wings but at a minor scale. 
    The latter can become relevant for sources showing high-amplitude variability, such as LPVs, where the change in PSF profile can impact amplitude measurements.

    \item The PSF model could also depend on the CCD location. 
    We tested a PSF model with additional dependency on the pixel and row position with respect to the center of the field of view and we did not find significant changes in PSF shape across the CCD.
    This is expected for Kepler observations, where the sky coverage of a single CCD ($\sim 1.2$ $deg^2$) is relatively small.
    Although, for larger fields of view instruments like the cameras in the Transiting Exoplanet Survey Satellite \citep[TESS, ][]{2015JATIS...1a4003R} that has CCDs covering $\sim 144 deg^2$ the PSF changes significantly within the CCD, making this model dependency necessary.

    \item The PSF model for CCD channels in the border of the field of view often exhibits extreme distortions and prominent features (see Figure 2 in \cite{2022AJ....163...93M} for a display of PSF models across CCDs) that could affect the model performance. 
    We tested light curves created with PSF models varying their flexibility (number of spline knots) and their center. 
    These alterations mimic possible miss-calculation of the centroids due to distorted PSF shapes and lack of model fidelity when steep gradients in the PSF profile are present. 
    We computed several metrics such as median flux, linear trend slope, amplitude, CDPP, and multiple flux percentile ratios, to assess the stability of the extracted light curves. 
    We found that even when the PSF centroid is missed by less than $2\arcsec$ (a half pixel) or when the model struggles with drastic gradient changes (e.g. a PSF shape with two close ``leg'' features), the light-curve metrics distributions are consistent between models. 
    This shows that our models are statistically robust to model parameters and small imperfections. 
    Although, some exceptions happen for highly blended and high-contrast sources. 
    The latter is the case for Tabby Star \citep[KIC 8462852, G = 11.6][]{2016MNRAS.457.3988B} and a contaminant fainter (G = 17.6) star (Gaia DR3 2081900944807842560) located less than $2 \arcsec$ away. 
    For both, our method produced imprecise photometry levels across quarters that were only possible to overcome when fitting the sources alone (i.e. removing one of them from the input catalog). 
    Although this approach is useful when working with a specific target, it is not optimal when performing massive source extraction.
    
    \item As described in section \ref{subsubsec:corr}, correlations can still be found between highly variable-bright sources and fainter neighbors, especially when they are close on the detector. 
    We computed a correlation metric across pairs of light curves and removed all faint sources that showed a correlation metric above the threshold. 
    Although this metric is effective in removing correlated sources, remaining small correlations can still be present leading to detecting the wrong cause of variability 
    We encourage users of these light curves to further analyze neighbor sources to secure the true origin of the variability.

    \item Although the perturbation model removes most of the velocity aberration and focus change trends, because the model is fitted for the entire scene these trends are not fully suppressed in every target. 
    This is particularly true for highly blended sources and with partial data where only the wings of the PSF are used to fit the models.
    \item The LFD photometry method relies on solving a linear model by means of least-square minimization. 
    This simplifies and enables rapid model fitting and light curve extraction, but no physical constraint are placed on the expected flux values, therefore negative solutions are mathematically possible. 
    We mitigate this issue by iterating the solving step while narrowing the priors (see \ref{subsec:priors}) for sources with predicted negative fluxes and their contaminating neighbors (which force the negative solution). 
    Sources that still have negative fluxes after the iteration is completed are rejected from the final catalog. 
    We found this affect mainly sources with mid- to high-contrast ($\ge 2$ mag) and within $15 \arcsec$.

\end{enumerate}

\begin{figure}[htb!]
    \centering
    \epsscale{1.15}
    \plotone{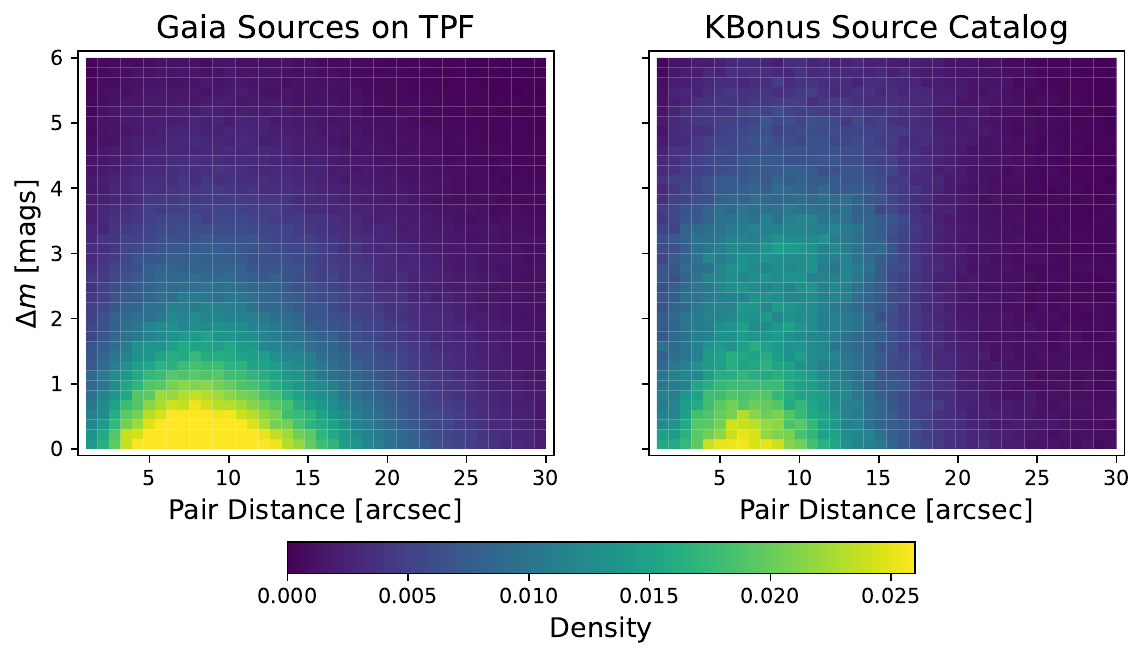}
    \caption{@d Histogram of magnitude contrast as a function distance for pair sources. The left panel shows the density of all the available Gaia sources ($G \leq 19$) on the TPFs which is our input catalog. The right panel shows the density of extracted sources in the KBonus catalog presented in this work. This figure shows the extraction biases discussed in Section~\ref{sec:limitations} in (1), (2), (4), and (7).}
    \label{fig:contrast_v_dist}
\end{figure}

Figure \ref{fig:contrast_v_dist} illustrates the combined extraction biases discussed above in (1), (2), (4), and (7). 
The figure shows the magnitude contrast and distance distribution for pairs of sources from the input and extracted catalogs. 
The former is denser for low-contrast and blends between 5 and 13 $\arcsec$ mainly because of sources slightly outside the TPF that get rejected due to insufficient pixel data (2) and fewer due to predicted negative fluxes (7). 
The decrease in density beyond 15 $\arcsec$ seen in the extracted catalog is due to the typical size of TPFs. 
The absence of pairs within $1 \arcsec$ is due to the selection bias described in (1). 
The apparent larger number counts in the extracted catalog for pairs with mid- to high-contrast is mainly due to (1) and (2) and partially to (4).


\section{Future Work} \label{sec:future_work}

A natural step forward is to use the \texttt{psfmachine} library to extract light curves from K2 \citep{2014PASP..126..398H} mission and the TESS mission. 

K2 data presents one major challenge, the failure of the spacecraft's reaction wheels caused a loss in the telescope pointing precision leading to a strong and characteristic jitter motion with a half-day timescale. 
This jitter motion drastically affects our perturbed model. 
First, the scene motion is not smooth anymore and the binning done to fit the perturbed model needs to increase in resolution to capture the motion. 
Increasing the time resolution of the perturbed model increases memory usage and computing time. 
Secondly, the CBV vectors is likely not be the best basis vectors to fit the perturbed model. 
Preliminary results have shown that using the centroid (or the mission positional corrector \texttt{pos\_corr}) vectors leads to better corrected light curves. 
An alternative approach is to compute PSF models and offset corrections for every cadence. 
Fitting a PSF model per cadence requires a large number of objects and pixels available, which can only be achieved by increasing the number of TPFs or when working with K2 superstamps pixel masks \citep{2018RNAAS...2Q..25C}, adding to computing costs.

By its design, the LFD method and the Python implementation \texttt{psfmachine} work well with TESS data after fine-tuning model parameters that account for the difference in pixel scale (TESS is $21\arcsec$/pix), integration times, and crowding effects. 
Although it is tentative to compile large catalogs of light curves for the entire TESS archive (several TB of TPF data), we believe that providing the users with a well-build and robust Python library able to quickly extract light curves (by using pre-computed models) or with full control of model parameters, represents a bigger contribution to the community.
Moreover, there are other active pipelines extracting similar PSF photometry to TESS primary and extended mission data.  
\cite{Han_2023} follows a similar approach using Gaia DR3 as input catalog and fits the effective PSF and background signal as a single linear model. 
This model is fitted later to every source but the extracted target to create a model of the full image, subtract this model from the data and then perform photometry on the decontaminated image of the target source. 
This approach is limited by the assumption that the background level is constant at the target's location and that stars around it are constant.
By design, the LFD method does not assume this and could improve on light curve precision. 

The current state of the \texttt{psfmachine} API implements loading PSF profile models pre-computed from Kepler's FFIs as discussed in Section \ref{subsec:PSF_model}. 
This enables users to quickly perform PSF photometry on single sources or on a small number of TPFs. 
However, this is only limited to the use of the mean-PSF model and not the full perturbed PSF model. 
This limitation is suitable for Kepler data where the perturbed-PSF model can only be fitted with a moderate number ($\geqslant 150$) of TPFs and not with the FFIs due to the low number of cadences per quarter. 
TESS FFIs are observed with a 30- or 10-minute cadence. 
This data presents the opportunity to compute and save the perturbed model for posterior extraction of light curves from any TESS data.
We plan to extend the \texttt{psfmachine} API to implement the saving and loading of the perturbed PSF model, resulting in a way to extract time-series from individual TPFs, using our best fit perturbation model. 
These new methods will speed up the photometry extraction using the fully corrected model, especially when extracting a small number of targets.

Light curve extraction from TESS FFIs is also possible with \texttt{psfmachine}, with the caveat that this process is considerably more memory intensive due to the loading of thousands of 2048 x 2048 pixel images.
A tractable solution is the combination of processing the FFIs in small cutouts (e.g. a 200 x 200 pixels cut has sufficient sources to estimate robust PSF models) and using pre-computed models.


\section{Summary} \label{sec:summary}

Kepler's primary mission consisted of eighteen 90-day quarters during which the telescope constantly observed the same field of view for almost 4 years.
These observations enable the community to find thousands of new exoplanets using the transit method, as well as perform numerous stellar variability and transient studies.
The Kepler mission delivered more than 200,000 image cutouts around previously selected targets and their aperture photometry light curves.
In this work, we reanalyze the image cutouts and extract PSF photometry light curves for all sources detected in the pixel data.
We created a catalog with 606,900 extracted sources from which 406,548 are new light curves from background sources. 
These background sources are objects detected on the pixel data but do not correspond to Kepler targets.
In our extraction pipeline, we used the method described in the LFD photometry method \citep{2021AJ....162..107H}.
The LFD method performs PSF photometry in a collection of TPFs by modeling the scene simultaneously.
It leverages the accuracy and precision of Gaia catalogs to fix the source locations and estimate a PSF model of the scene.
The method also computes corrections to the PSF model to account for the scene motion due to the velocity aberration effect, focus change, and pointing instabilities. 
Our extraction pipeline includes background modeling and subtraction, PSF fitting and photometry, aperture photometry using the PSF profile shape, and numerous extraction metrics useful to characterize the quality of the data.
The light curves produced in this work are available for public access via the MAST archive.
We implemented new methods and routines to the Python package \texttt{psfmachine} such as the background modeling, PSF model loading from pre-computed ones using FFIdata, and user-defined basis vectors for the perturbation model.
These new features are included in v1.1.4 of \texttt{psfmachine}.

We demonstrated that the quality of our light curves reaches similar accuracy levels as those delivered by the Kepler pipeline. 
The computed CDPP values range from 10s ppm for sources brighter than G=14 to 100s ppm for sources between 16th and 18th magnitude. 
Statistically, PSF photometry performs up to 40\% better, in CDPP value, compared to aperture photometry for sources fainter than G=13.25.
We listed and discussed the limitations of our extraction pipeline and the resulting light curves.
This serves as guidelines for users of this dataset.

We show two applications as examples of what can be accomplished with these high-level science products.
First, we compared the transit depth and estimated exoplanet radius between PDCSAP and our PSF light curves. 
The result suggests that the PSF photometry yields slightly deeper transits and therefore larger planets.
Although, we did not find significant changes to the planet size-period relationship.
Secondly, we show examples of the power of PSF photometry to deblend contaminated sources by revisiting KOI false positives due to background binary contamination.
The LFD photometry method successfully separates highly blended sources at high contrast which is relevant to distinguish false positives from real exoplanet candidates.
This new dataset presents other numerous opportunities to the community not limited only to exoplanet studies.
To name some, there are 50 new white dwarf light curves which add to the original 41 in the Kepler target list, thousands of light curves of potential M-dwarf stars, and expand asteroseismic analysis of rotating stars.


\begin{acknowledgments}

This paper includes data collected by the Kepler mission and obtained from the MAST data archive at the Space Telescope Science Institute (STScI). Funding for the Kepler mission is provided by the NASA Science Mission Directorate. STScI is operated by the Association of Universities for Research in Astronomy, Inc., under NASA contract NAS 5–26555. 
This work has made use of data from the European Space Agency (ESA) mission Gaia (\url{https://www. cosmos.esa.int/gaia}), processed by the Gaia Data Processing and Analysis Consortium (DPAC, \url{https://www.cosmos.esa. int/web/gaia/dpac/consortium}). Funding for the DPAC has been provided by national institutions, in particular, the institutions participating in the Gaia Multilateral Agreement. 
Resources supporting this work were provided by the NASA High-End Computing (HEC) Program through the NASA Advanced Supercomputing (NAS) Division at Ames Research Center.
Funding for this work for JMP is provided by grant number 80NSSC20K0874, through NASA ROSES.

\end{acknowledgments}

\vspace{5mm}
\facilities{Kepler}

\software{astropy \citep{2013A&A...558A..33A},  
          lightkurve \citep{2018ascl.soft12013L}, 
          numpy \citep{harris2020array},
          psfmachine \citep{2021zndo...4784073H},
          scipy \citep{2020SciPy-NMeth},
          }

\appendix

\restartappendixnumbering

\section{KBonus Light Curve Examples: Confirmed Planets} \label{appx:lc_ex_exo}

Figures \ref{fig:exo_1}, \ref{fig:exo_2}, \ref{fig:exo_3}, \ref{fig:exo_4}, and \ref{fig:exo_5} compare Kepler PDCSAP and our PSF light curves for confirmed exoplanets at different signal-to-noise ratio levels.

\begin{figure*}[htb!]
    \centering
    \epsscale{1.15}
    \plotone{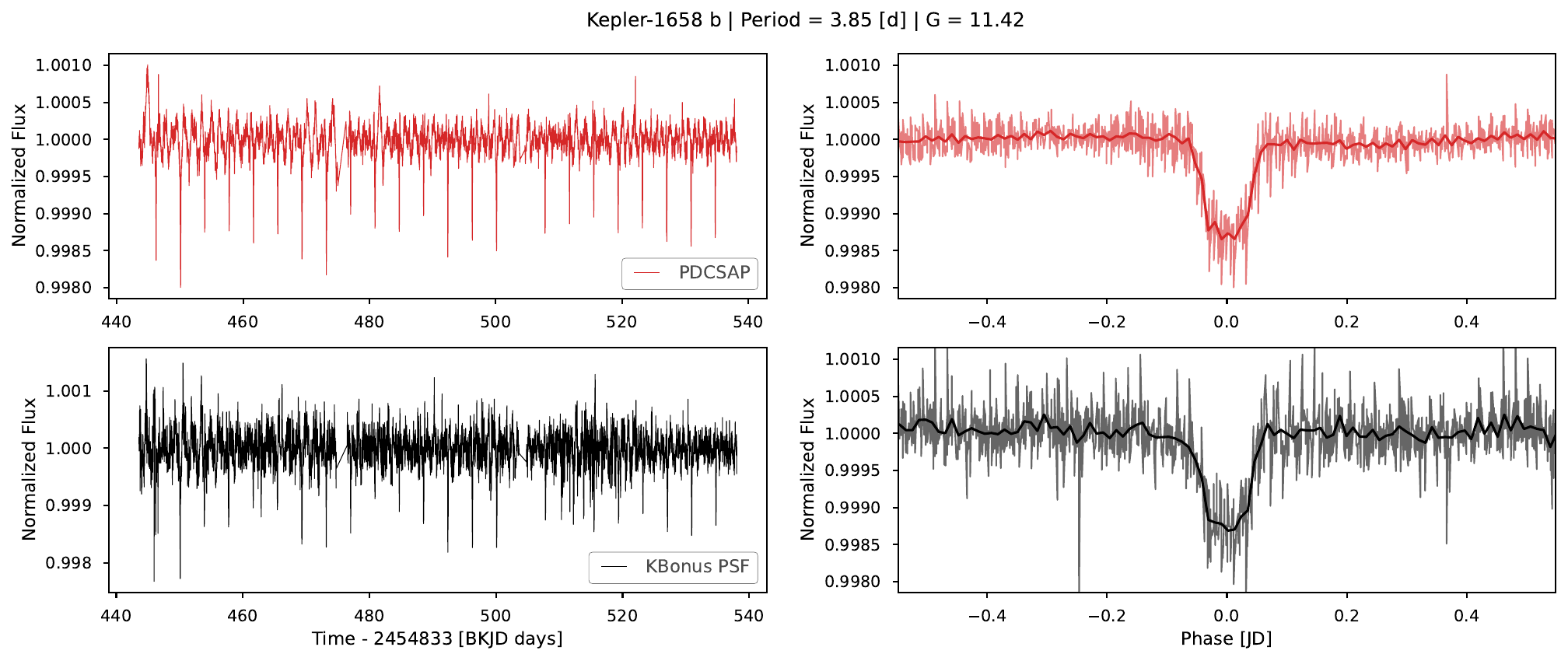}
    \caption{Confirmed exoplanet Kepler-1698 b with an orbital period of 3.85 days. The top panels (red) show Kepler's PDCSAP light curve, while the bottom panels (black) are the KBonus PSF photometry. The left panels show the light curve in time space and the right panels show the phase-folded light curve centered in mid-transit.}
    \label{fig:exo_1}
\end{figure*}

\begin{figure*}[htb!]
    \centering
    \epsscale{1.15}
    \plotone{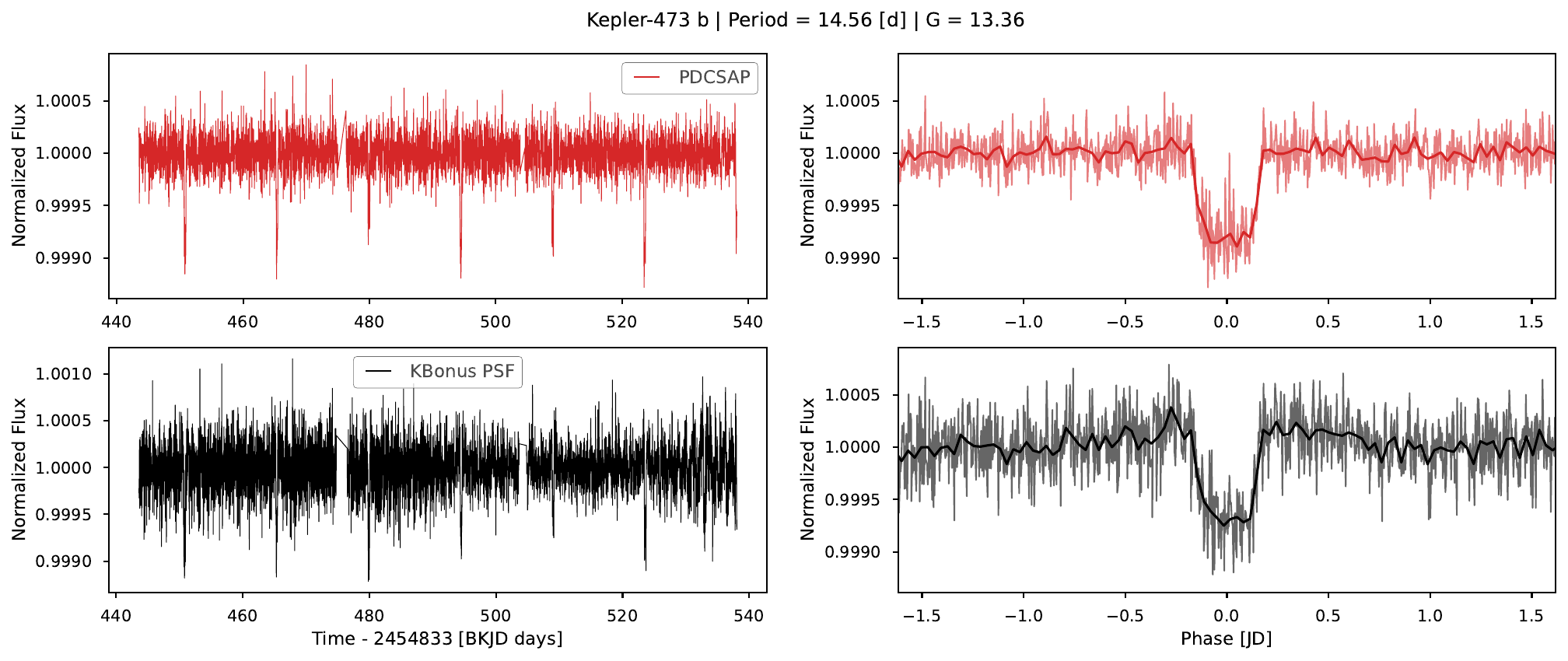}
    \caption{Similar to Figure \ref{fig:exo_1} for confirmed exoplanet Kepler-473 b with an orbital period of 14.56 days.}
    \label{fig:exo_2}
\end{figure*}

\begin{figure*}[htb!]
    \centering
    \epsscale{1.12}
    \plotone{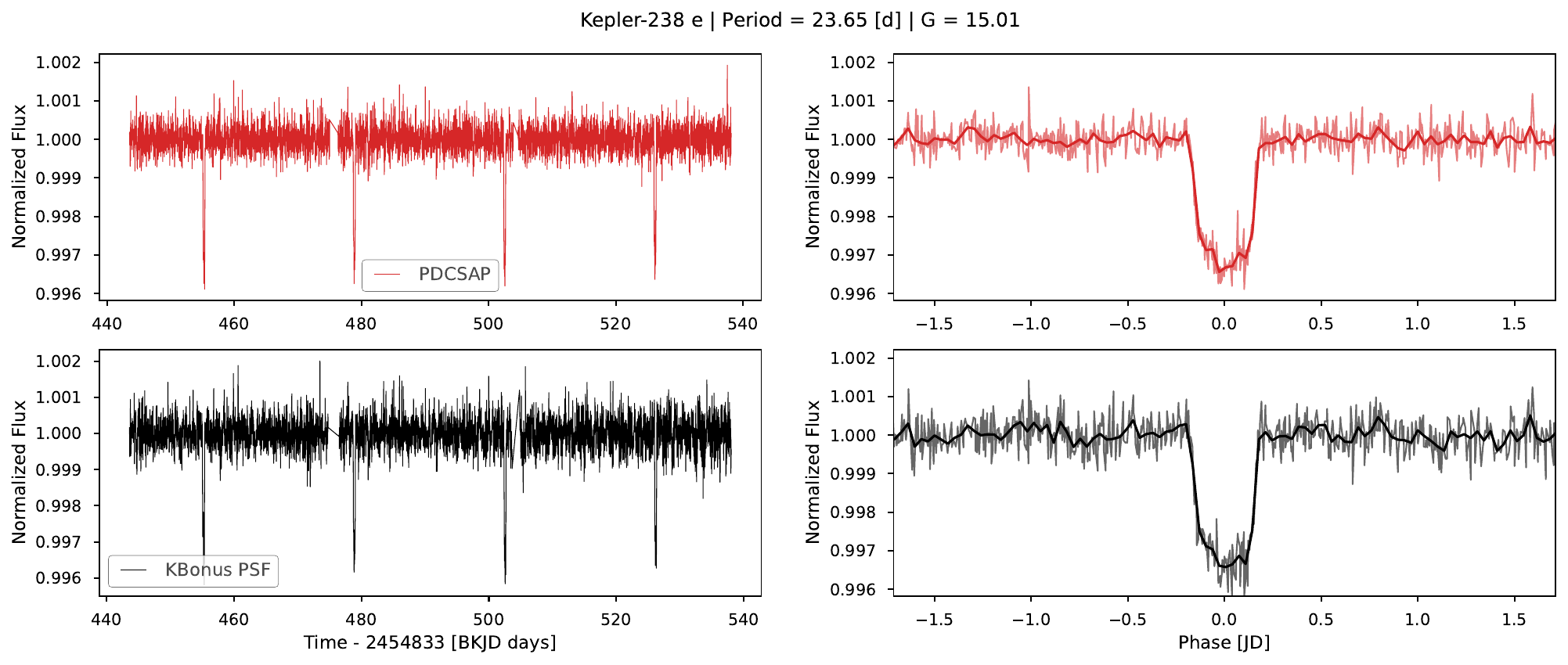}
    \plotone{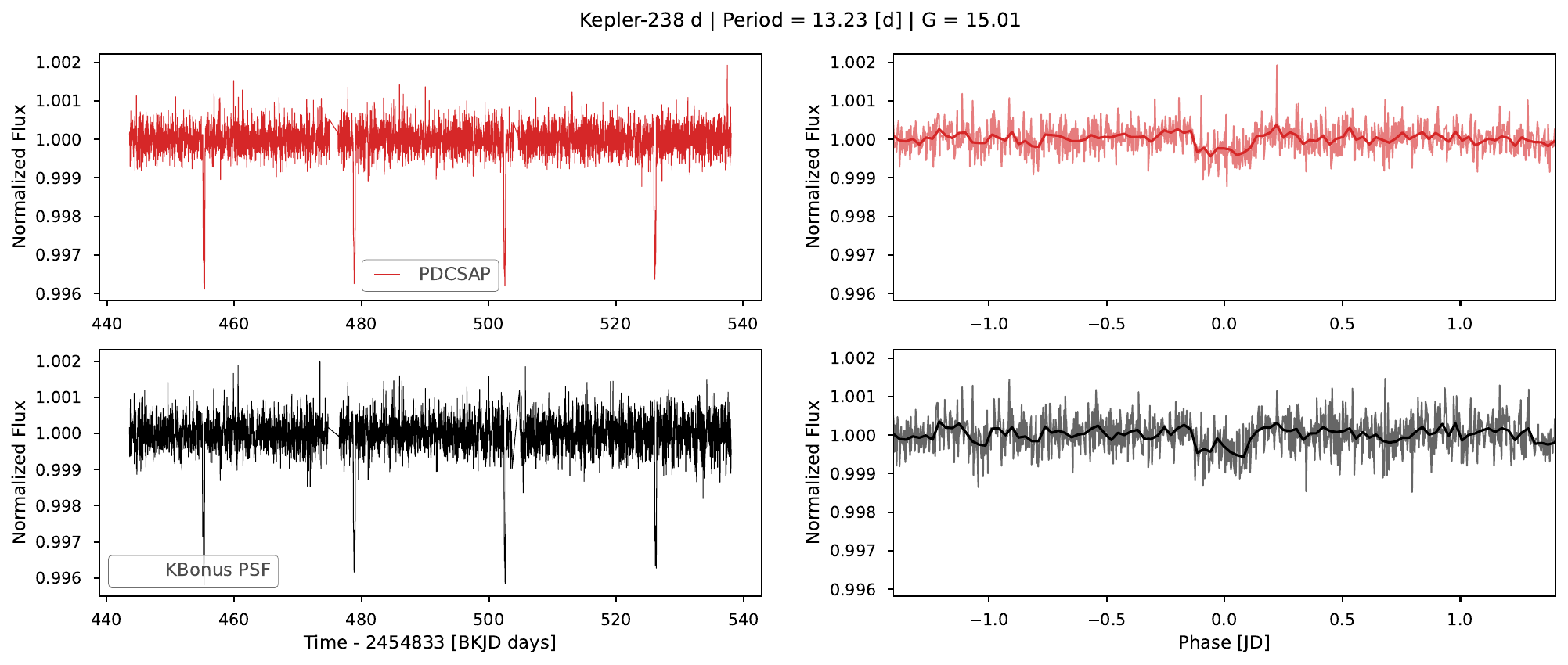}
    \plotone{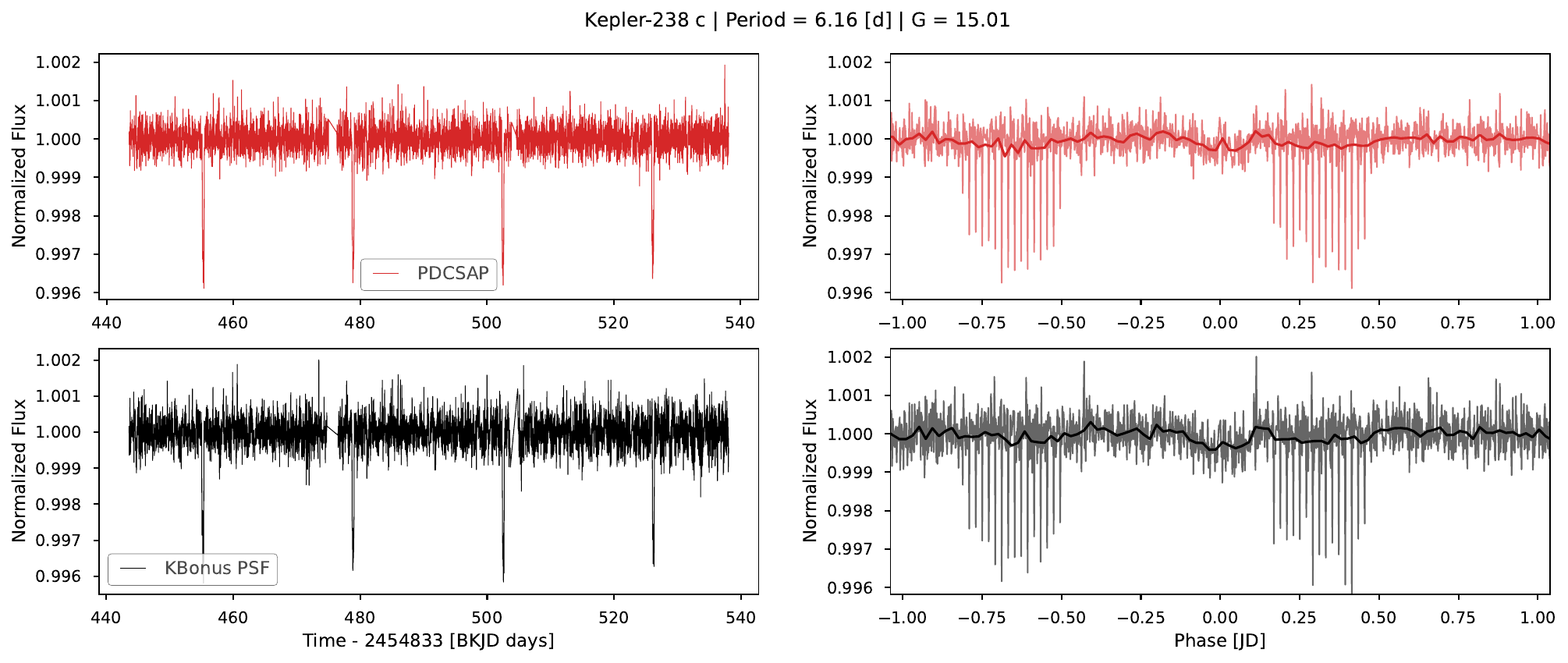}
    \caption{Confirmed multi-planet system Kepler-238 e, Kepler-238 d, and Kepler-238 c, with orbital periods 23.65, 13.23, and 6.16 days respectively. Similar to Figure \ref{fig:exo_1}.}
    \label{fig:exo_3}
\end{figure*}

\begin{figure*}[htb!]
    \centering
    \epsscale{1.15}
    \plotone{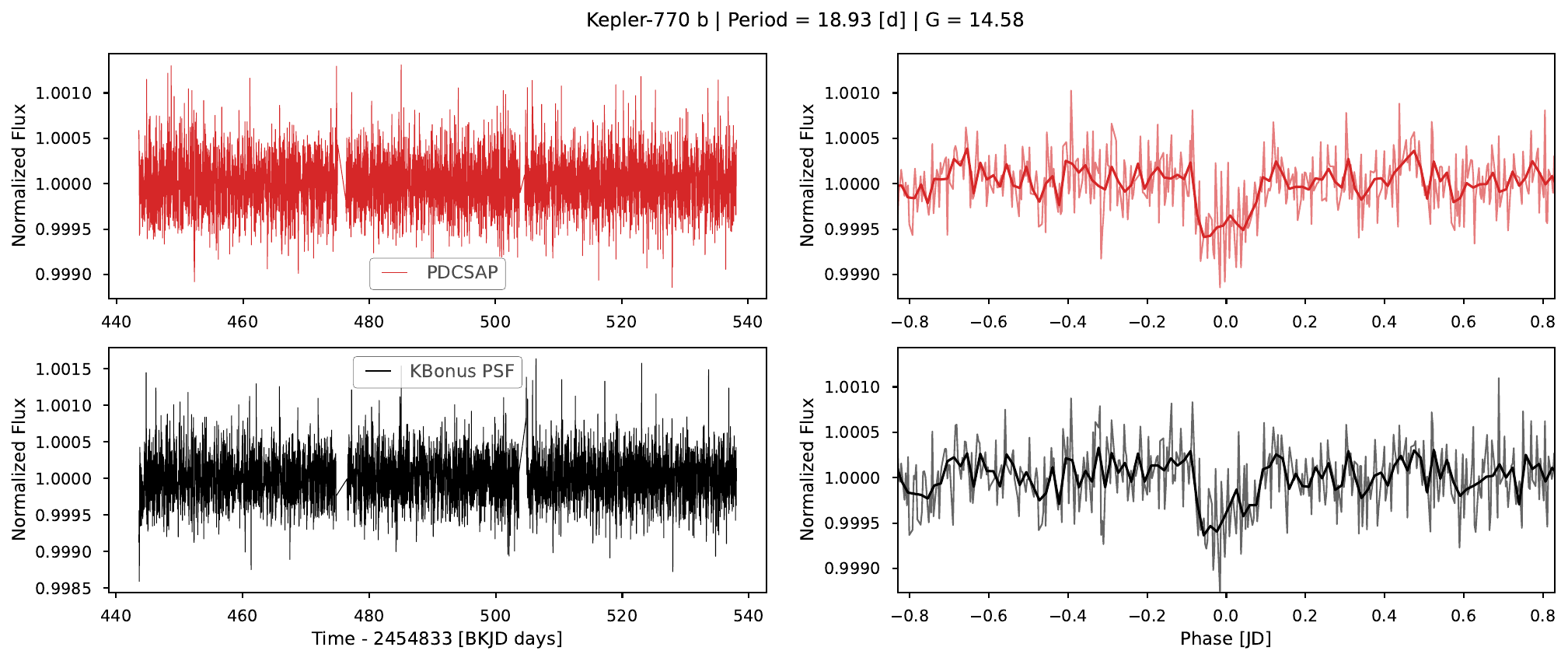}
    \caption{Confirmed exoplanet Kepler-770 b with an orbital period of 18.93 days. Similar to Figure \ref{fig:exo_1}.}
    \label{fig:exo_4}
\end{figure*}

\begin{figure*}[htb!]
    \centering
    \epsscale{1.15}
    \plotone{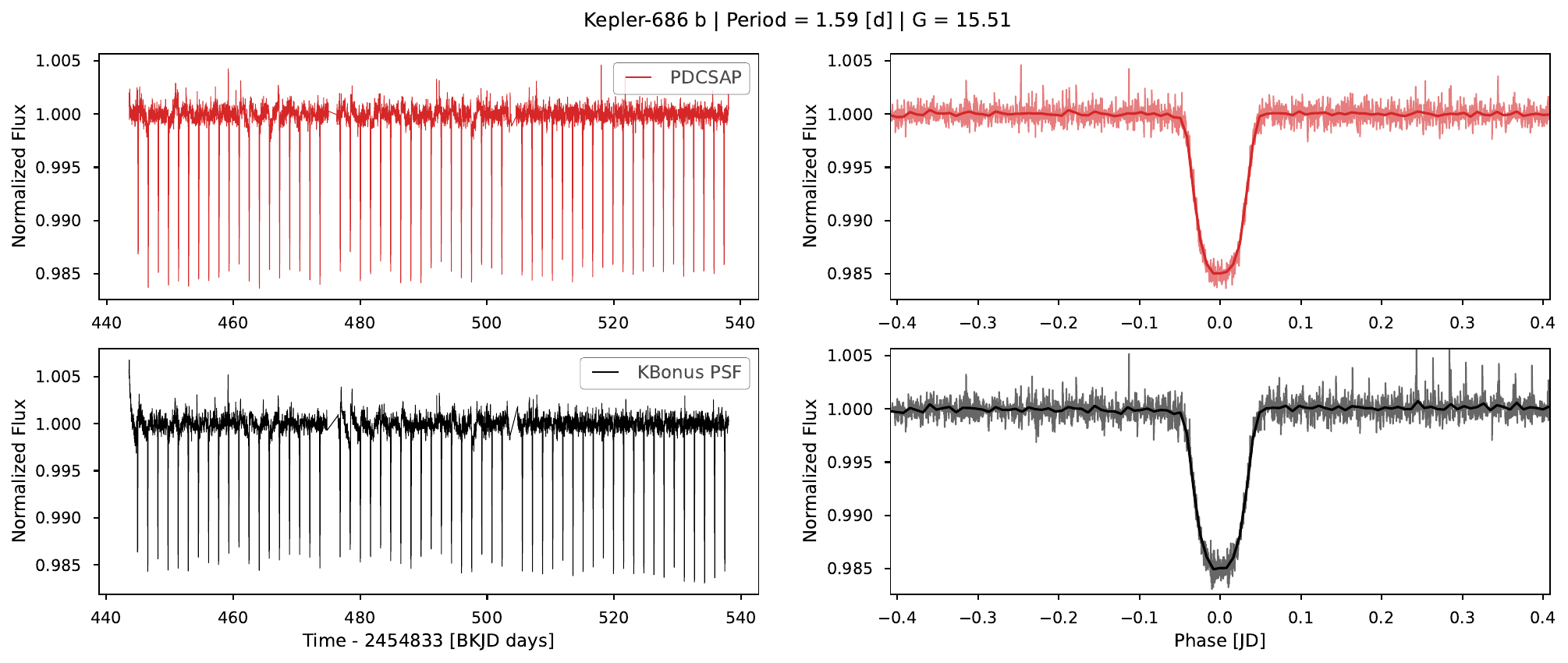}
    \caption{Confirmed exoplanet Kepler-686 b with an orbital period of 1.59 days. Similar to Figure \ref{fig:exo_1}.}
    \label{fig:exo_5}
\end{figure*}

\restartappendixnumbering

\section{KBonus Light Curve Examples: KOIs} \label{appx:lc_ex_koi}

Figures \ref{fig:koi_3}, \ref{fig:koi_4} and \ref{fig:koi_5} complement Section \ref{subsec:kois} showing three more false positive KOI examples.

\begin{figure*}[htb!]
    \centering
    \epsscale{1.15}
    \plotone{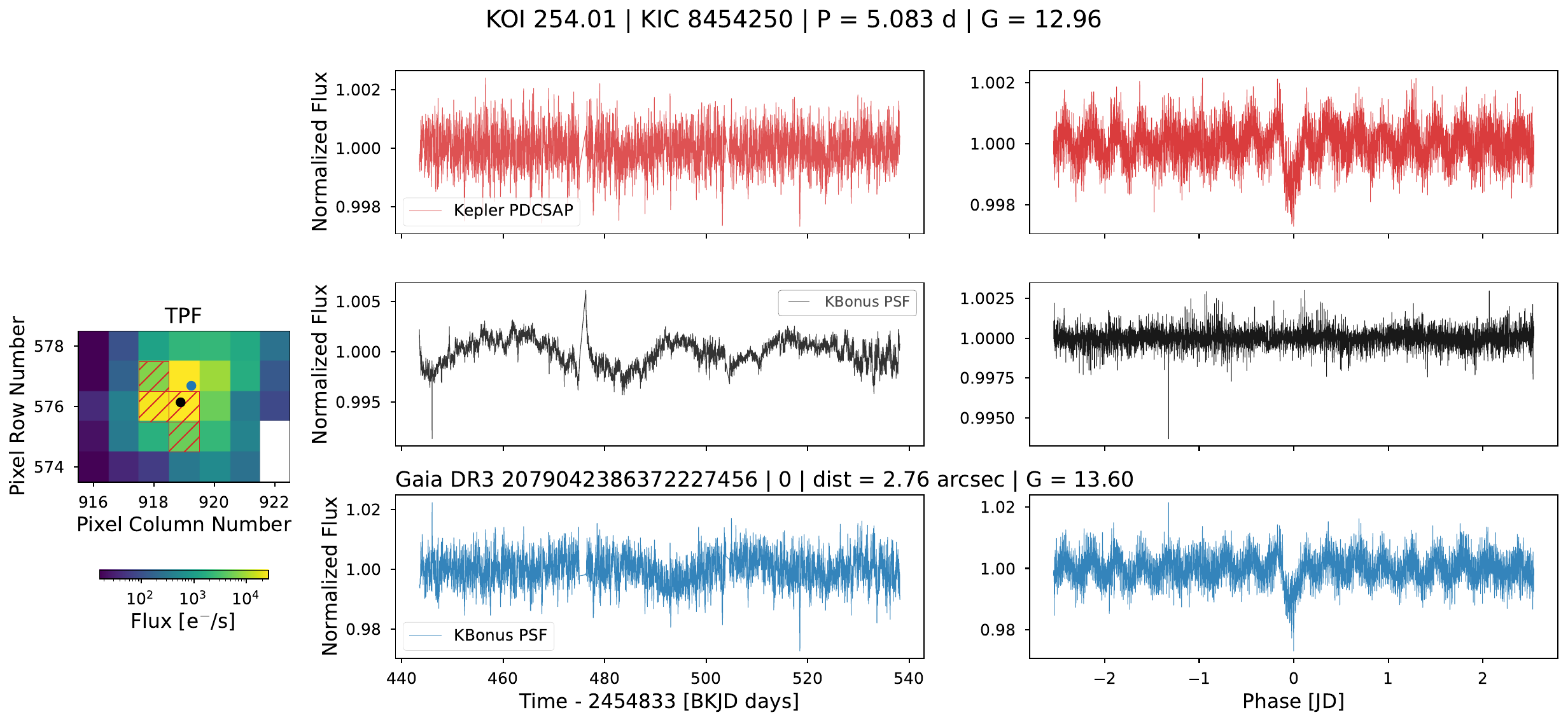}
    \caption{KOI 254.01 false positive candidate due to centroid offset. Figure layout and legends are the same as Figure \ref{fig:koi_1}.}
    \label{fig:koi_3}
\end{figure*}

\begin{figure*}[htb!]
    \centering
    \epsscale{1.15}
    \plotone{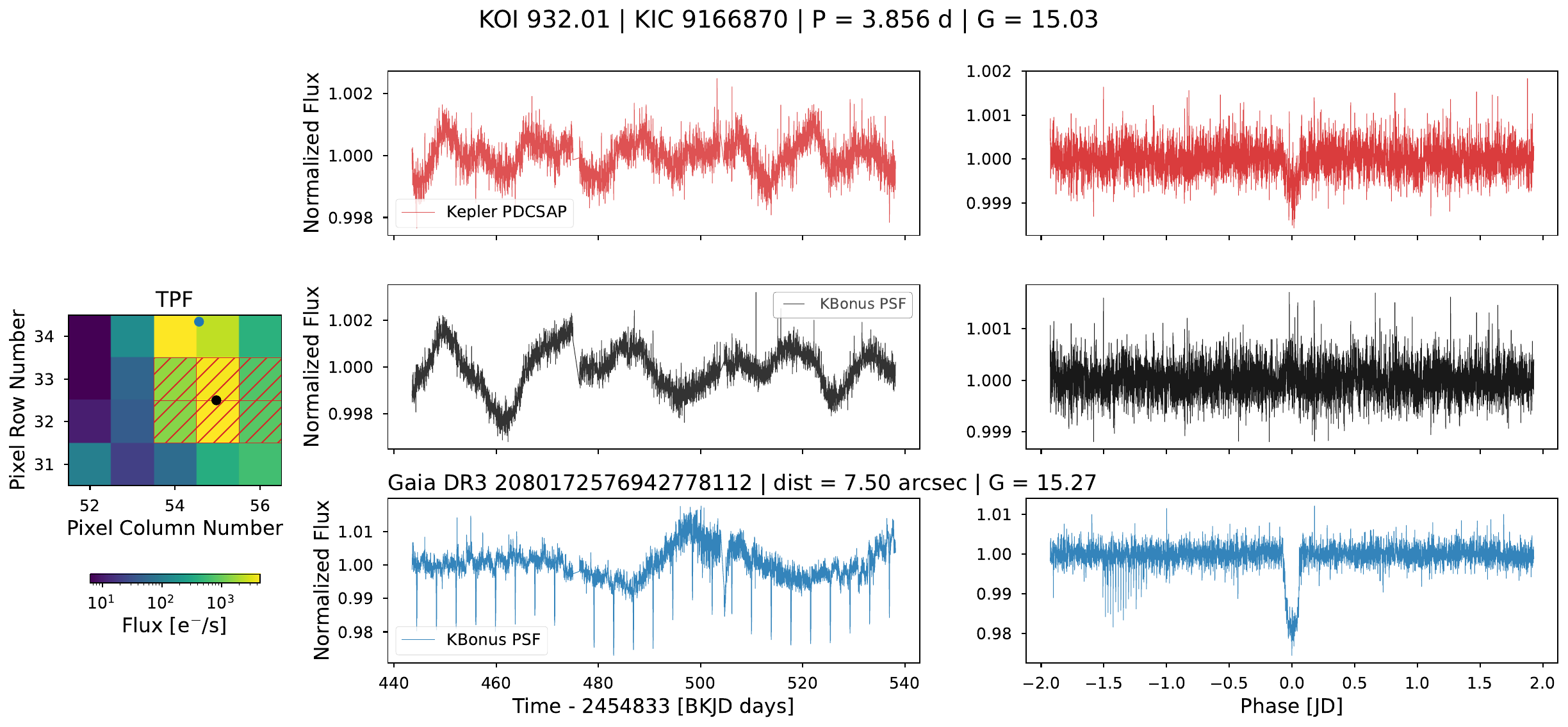}
    \caption{KOI 932.01 false positive candidate due to centroid offset. Figure layout and legends are the same as Figure \ref{fig:koi_1}. The contaminant is KIC 9166862, which hosts the confirmed planet Kepler-731 b.}
    \label{fig:koi_4}
\end{figure*}

\begin{figure*}[htb!]
    \centering
    \epsscale{1.15}
    \plotone{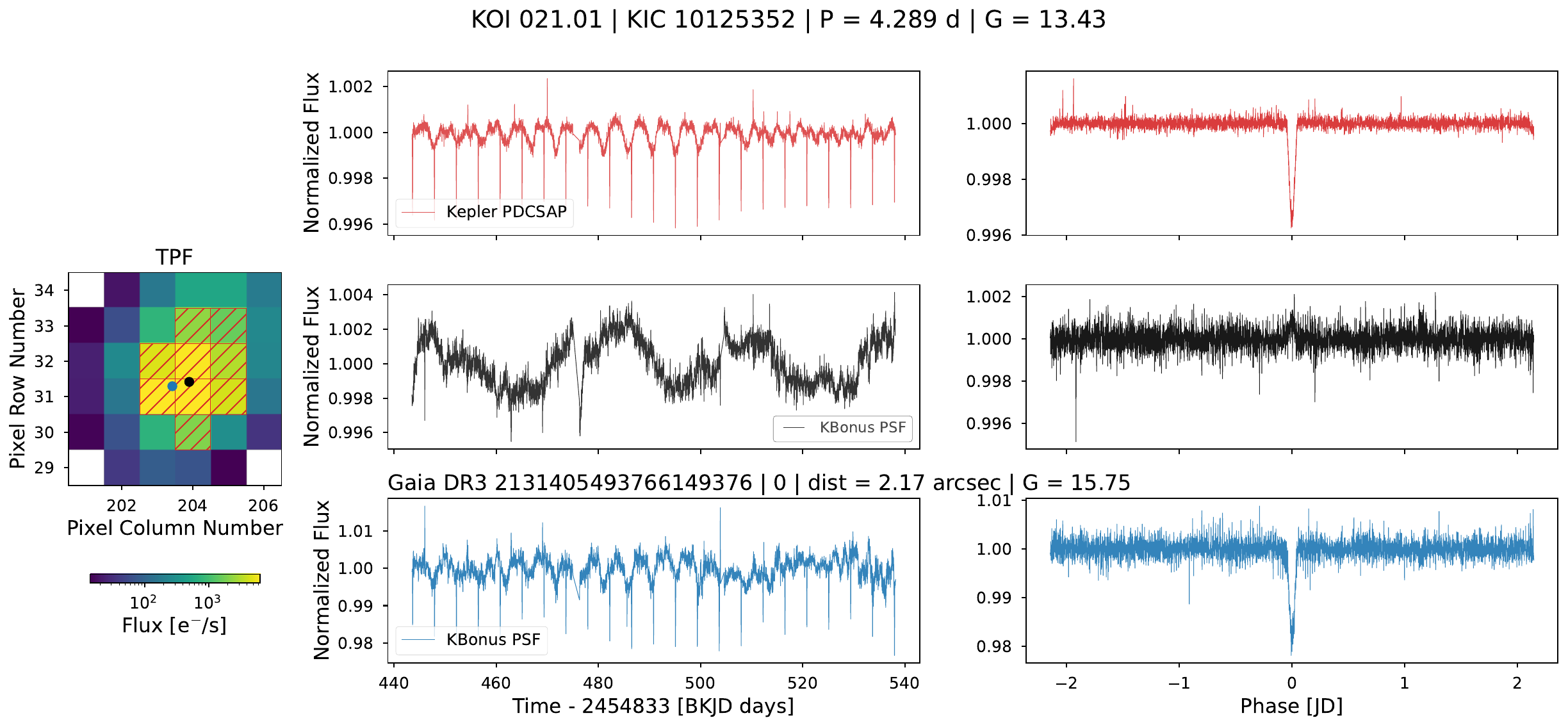}
    \caption{KOI 021.01 false positive candidate due to centroid offset. Figure layout and legends are the same as Figure \ref{fig:koi_1}.}
    \label{fig:koi_5}
\end{figure*}

\restartappendixnumbering

\section{KBonus Light Curve Examples: Long Period Variables} \label{appx:lc_ex_lpv}

We selected ten LPV from the Gaia DR3 variable catalog \citep{2022arXiv220605745L} to illustrate the inter-quarter photometry. 
Figures \ref{fig:lpvs_1} and \ref{fig:lpvs_2} show the light curve examples and the PDCSAP photometry for comparison.
Thanks to the perturbed model (see Section \ref{subsec:per_model} that fits the scene velocity aberration and long-term instrumental trends, stellar long-term variability is preserved, and the photometry from consecutive quarters matches in almost all cases.

\begin{figure*}[htb!]
    \centering
    \plotone{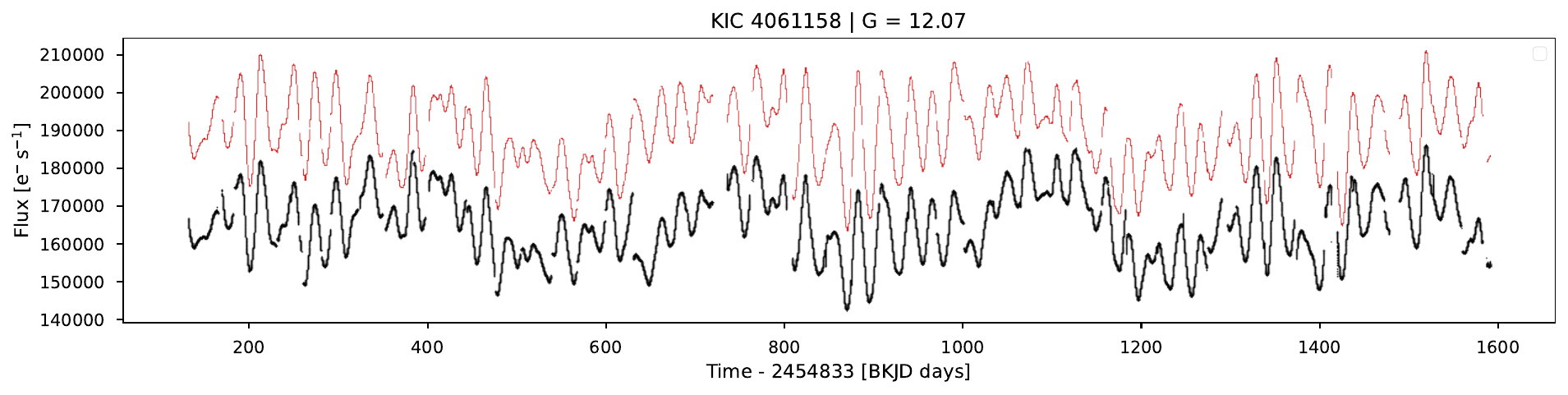}
    \plotone{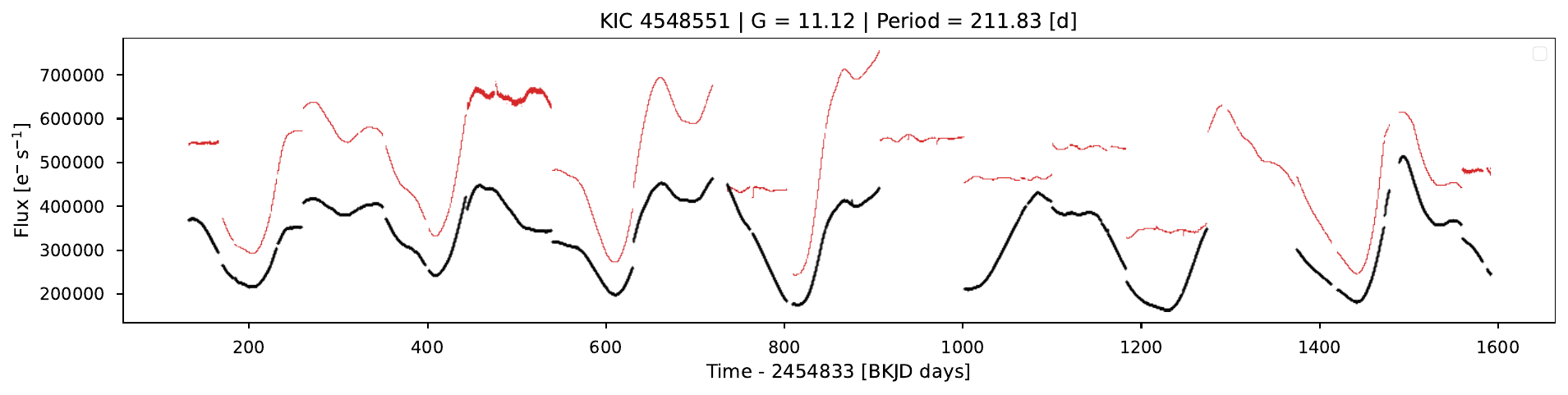}
    \plotone{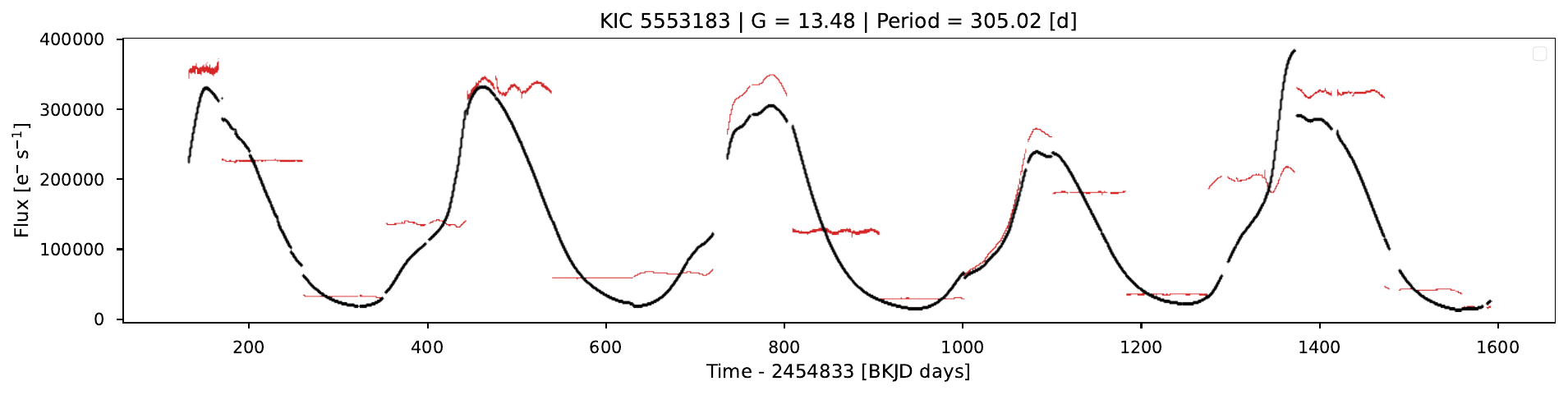}
    \plotone{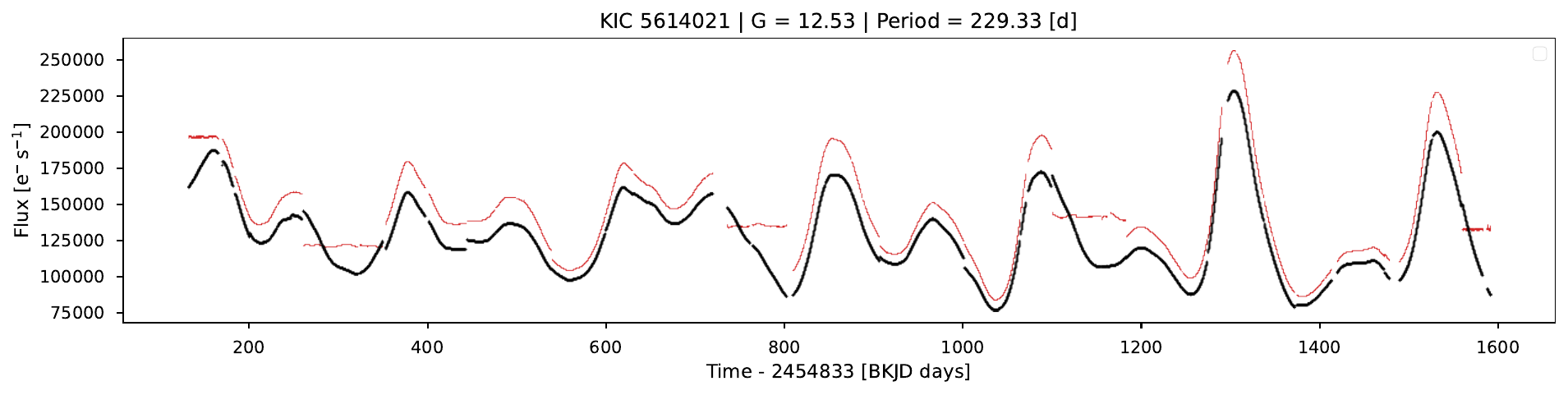}
    \plotone{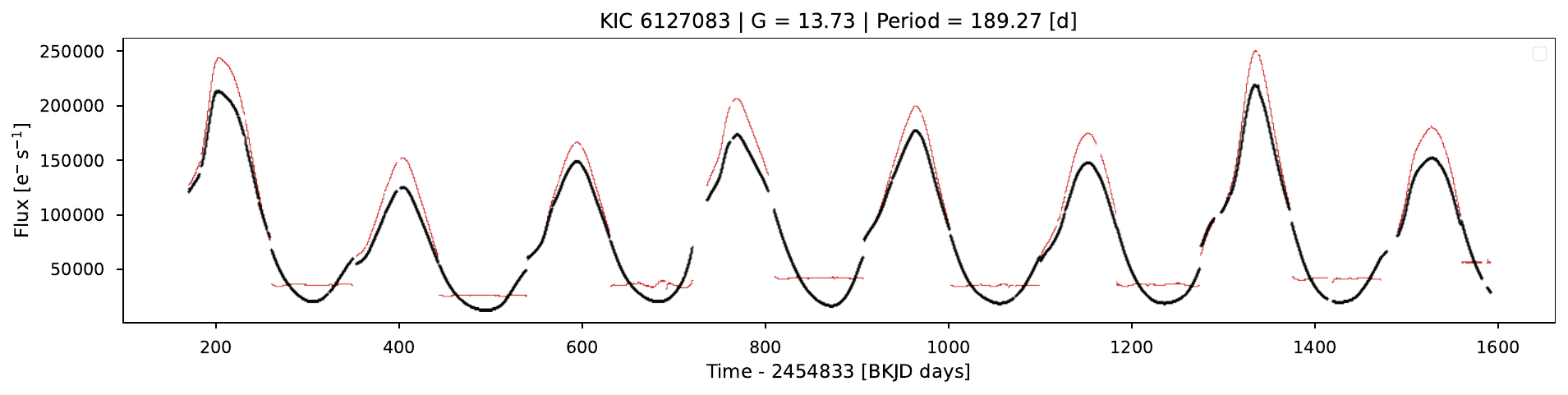}
    \caption{Five LPV light curve examples selected from the Gaia DR3 variable catalog. Red light curves are PDCSAP photometry, while black light curves are our PSF photometry. KIC identifier, G-band magnitude, and reference periods (when available) are in the title of each panel. Missing quarter data is due to non-detection, or subpar PSF estimation (e.g. PSF fraction below 0.5 threshold).}
    \label{fig:lpvs_1}
\end{figure*}

\begin{figure*}[htb!]
    \centering
    \plotone{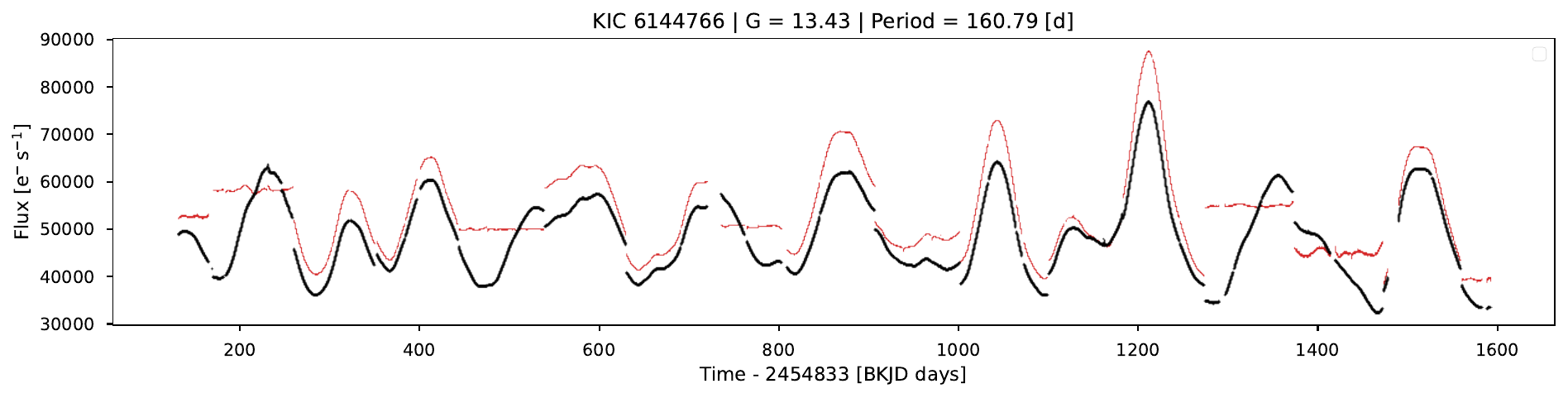}
    \plotone{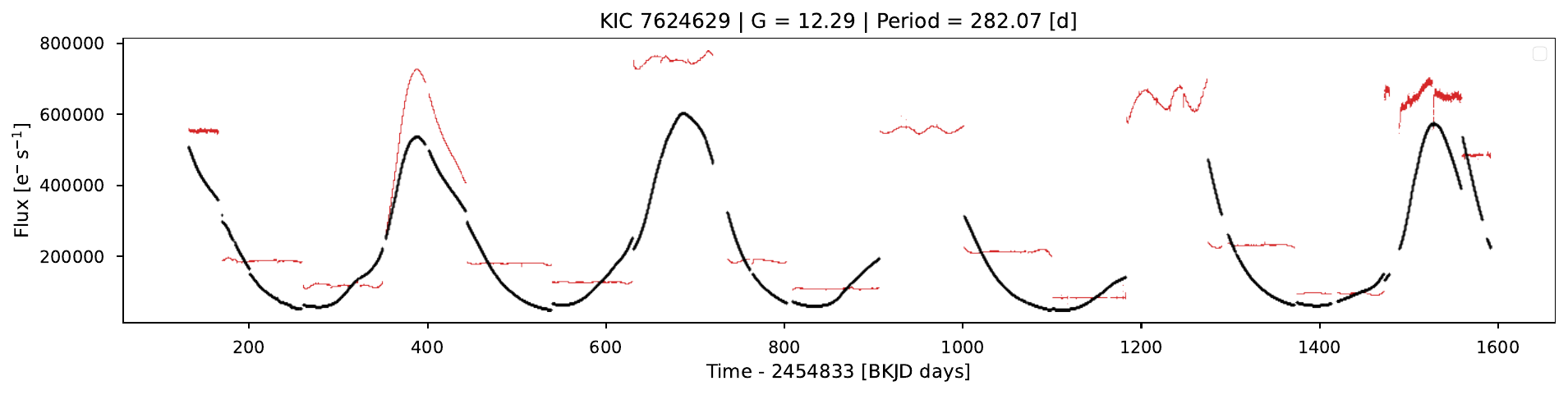}
    \plotone{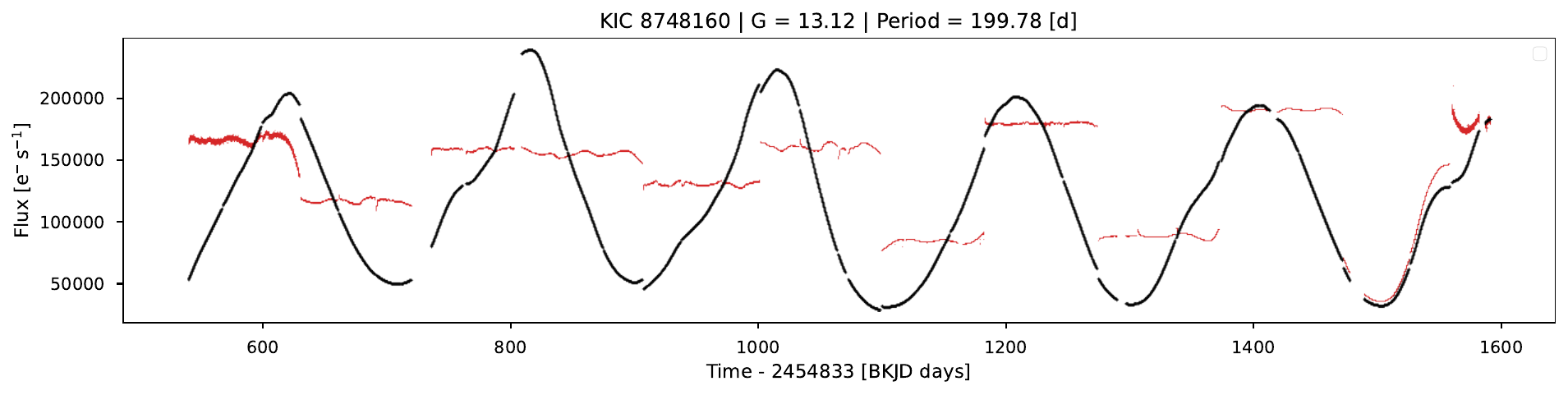}
    \plotone{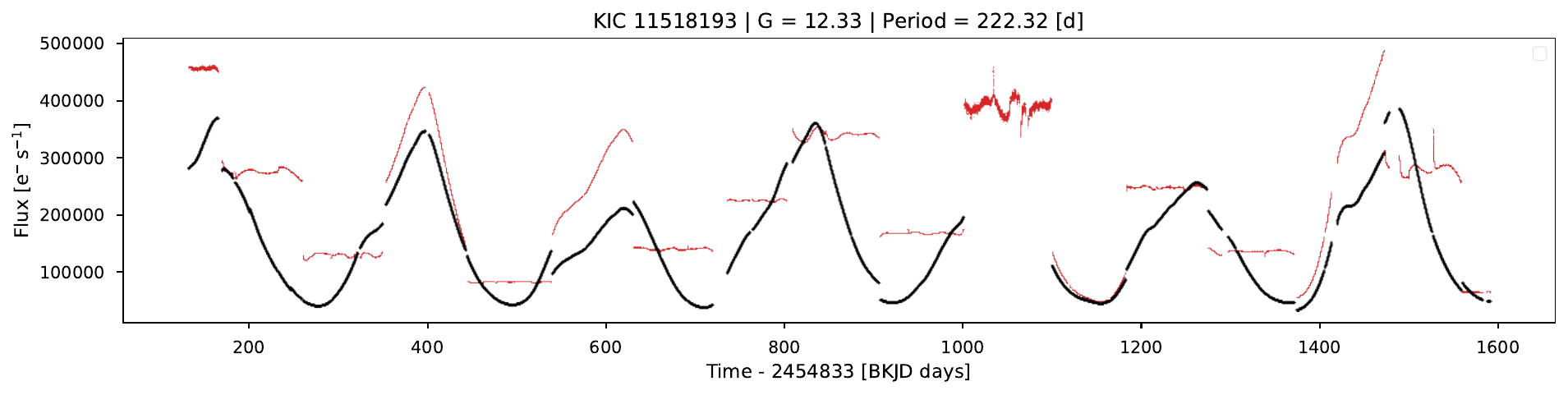}
    \plotone{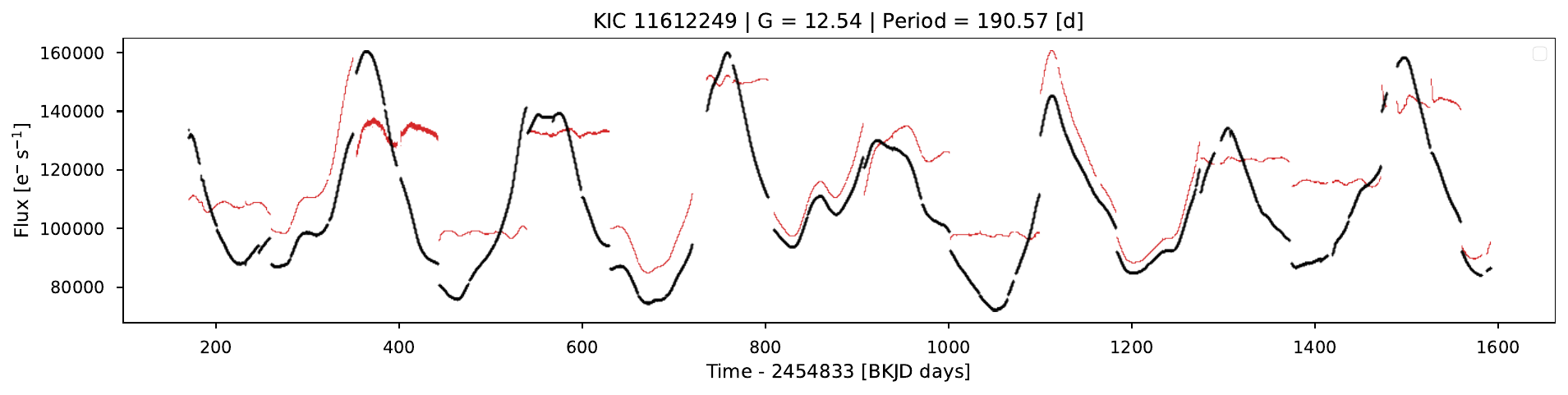}
    \caption{Next five LPV light curve examples. Similar to Figure \ref{fig:lpvs_1}.}
    \label{fig:lpvs_2}
\end{figure*}

\restartappendixnumbering

\section{KBonus Light Curve Files} \label{appx:lcfs}

The light curve files are delivered as multi-extension FITS files following a similar organization as the original Kepler LCFs. 
Each file is named with the following pattern \texttt{hlsp\_kbonus-bkg\_kepler\_kepler\_<source\_id>\_\linebreak kepler\_v1.0\_lc.fits} where \texttt{source\_id} is the corresponding Kepler Input Catalog number (e.g. \texttt{kic\_005018361}) if the source exists in the catalog or the Gaia DR3 number(e.g. \texttt{gaia\_dr3\_2077321719392686592}) if not. 
Table \ref{tab:lcf_fits} shows a description of the FITS files. The LIGHTCURVE{\_}STITCHED is the main extension containing the fully stitched time series using quarters with a PSF fraction greater than 0.5. 
Table \ref{tab:stitch_lc} details the column in this extension. The LIGHTCURVE{\_}Q extension contains the light curve for single quarters, Table \ref{tab:q_lc} details its content, this extension contains per-quarter metrics and other measurements such as centroid values. 
The APERTURE{\_}Q extension contains the pixel mask of the corresponding quarter used for the aperture photometry in the shape of the TPF of origin. 
The FITS files only have LIGHTCURVE and APERTURE extensions for quarters where the source was detected, therefore the number of extensions varies. 

The FITS files are structured to work seamlessly with the \texttt{lightkurve}\citep{2018ascl.soft12013L} package. 
In this way, users can easily load the stitched light curve into a \texttt{Light Curve} object. 
See the KBonus documentation\footnote{\url{https://github.com/jorgemarpa/KBonus/tree/main}} for further details on how to work with these files.

    

\begin{deluxetable*}{lllcl}[htb!]
\tablecaption{Multi-extension FITS file example for file \texttt{hlsp\_kbonus-bkg\_kepler\_kepler\_kic-005098770\_kepler\_v1.0\_lc.fits}.
\label{tab:lcf_fits}}
\tablewidth{0pt}
\tablehead{
\colhead{No.} & \colhead{Name} & \colhead{Type} & \colhead{Cards} & \colhead{Dimensions}
}
\startdata
0  & PRIMARY                & PrimaryHDU  & 38 & -             \\
1  & LIGHTCURVE{\_}STITCHED & BinTableHDU & 40 & 65276R x 11C  \\
2  & LIGHTCURVE{\_}Q0       & BinTableHDU & 62 & 469R x 13C    \\
3  & APERTURE{\_}Q0         & ImageHDU    & 10 & (9, 8)        \\
4  & LIGHTCURVE{\_}Q1       & BinTableHDU & 62 & 1624R x 13C   \\
5  & APERTURE{\_}Q1         & ImageHDU    & 10 & (8, 7)        \\
6  & LIGHTCURVE{\_}Q2       & BinTableHDU & 62 & 4081R x 13C   \\
7  & APERTURE{\_}Q2         & ImageHDU    & 10 & (7, 6)        \\
8  & LIGHTCURVE{\_}Q3       & BinTableHDU & 62 & 4135R x 13C   \\
9  & APERTURE{\_}Q3         & ImageHDU    & 10 & (6, 6)        \\
10 & LIGHTCURVE{\_}Q4       & BinTableHDU & 62 & 4110R x 13C   \\
11 & APERTURE{\_}Q4         & ImageHDU    & 10 & (6, 5)        \\
12 & LIGHTCURVE{\_}Q5       & BinTableHDU & 62 & 4487R x 13C   \\
13 & APERTURE{\_}Q5         & ImageHDU    & 10 & (6, 6)        \\
14 & LIGHTCURVE{\_}Q6       & BinTableHDU & 62 & 4272R x 13C   \\
15 & APERTURE{\_}Q6         & ImageHDU    & 10 & (7, 6)        \\
16 & LIGHTCURVE{\_}Q7       & BinTableHDU & 62 & 4227R x 13C   \\
17 & APERTURE{\_}Q7         & ImageHDU    & 10 & (6, 6)        \\
18 & LIGHTCURVE{\_}Q8       & BinTableHDU & 62 & 3107R x 13C   \\
19 & APERTURE{\_}Q8         & ImageHDU    & 10 & (6, 5)        \\
20 & LIGHTCURVE{\_}Q9       & BinTableHDU & 62 & 4653R x 13C   \\
21 & APERTURE{\_}Q9         & ImageHDU    & 10 & (6, 6)        \\
22 & LIGHTCURVE{\_}Q10      & BinTableHDU & 62 & 4442R x 13C   \\
23 & APERTURE{\_}Q10        & ImageHDU    & 10 & (7, 6)        \\
24 & LIGHTCURVE{\_}Q11      & BinTableHDU & 62 & 4474R x 13C   \\
25 & APERTURE{\_}Q11        & ImageHDU    & 10 & (6, 6)        \\
26 & LIGHTCURVE{\_}Q12      & BinTableHDU & 62 & 3885R x 13C   \\
27 & APERTURE{\_}Q12        & ImageHDU    & 10 & (6, 5)        \\
28 & LIGHTCURVE{\_}Q13      & BinTableHDU & 62 & 4243R x 13C   \\
29 & APERTURE{\_}Q13        & ImageHDU    & 10 & (6, 6)        \\
30 & LIGHTCURVE{\_}Q14      & BinTableHDU & 62 & 4270R x 13C   \\
31 & APERTURE{\_}Q14        & ImageHDU    & 10 & (7, 6)        \\
32 & LIGHTCURVE{\_}Q15      & BinTableHDU & 62 & 4367R x 13C   \\
33 & APERTURE{\_}Q15        & ImageHDU    & 10 & (6, 6)        \\
34 & LIGHTCURVE{\_}Q16      & BinTableHDU & 62 & 3535R x 13C   \\
35 & APERTURE{\_}Q16        & ImageHDU    & 10 & (6, 5)        \\
36 & LIGHTCURVE{\_}Q17      & BinTableHDU & 62 & 1289R x 13C   \\
37 & APERTURE{\_}Q17        & ImageHDU    & 10 & (6, 6)        \\
\enddata
\end{deluxetable*}

\begin{deluxetable*}{lllcl}[htb!]
\tablecaption{Description of the columns available in the LIGHTCURVE{\_}STITCHED extension.
\label{tab:stitch_lc}}
\tablewidth{0pt}
\tablehead{
\colhead{Column} & \colhead{Field} &\colhead{Format} &\colhead{Units} &\colhead{Description}
}
\startdata
1  & Time                       & float64  & BJD - 2454833 & Time value in BKJD. \\
2  & Cadenceno                  & int32    & -             & Cadence number. \\
3  & Flux                       & float64  & $e^-/s$       & PSF flux from stitched quarters. \\
4  & Flux{\_}err                & float64  & $e^-/s$       & PSF flux error from stitched quarters. \\
6  & SAP{\_}flux                & float64  & $e^-/s$       & SAP flux from stitched quarters. \\
7  & SAP{\_}flux{\_}err         & float64  & $e^-/s$       & SAP flux error from stitched quarters. \\
8  & PSF{\_}flat{\_}flux        & float64  & $e^-/s$       & PSF flux from stitched quarters after flattening. \\
9  & PSF{\_}flat{\_}flux{\_}err & float64  & $e^-/s$       & PSF flux error from stitched quarters after flattening. \\
10 & SAP{\_}quality             & int32    & -             & Quality flag from the TPF. \\
11 & Flatten{\_}mask            & int32    & -             & Quality flag from the flattening process. \\
\enddata
\end{deluxetable*}

\begin{deluxetable*}{lllcl}[htb!]
\tablecaption{Description of the columns available in the LIGHTCURVE{\_}Q extensions.
\label{tab:q_lc}}
\tablewidth{0pt}
\tablehead{
\colhead{Column} & \colhead{Field} &\colhead{Format} &\colhead{Units} &\colhead{Description}
}
\startdata
1  & Time                       & float64  & BJD - 2454833 & Time value in BKJD. \\
2  & Cadenceno                  & int32    & -             & Cadence number. \\
3  & Flux                       & float64  & $e^-/s$       & Corrected PSF flux. \\
4  & Flux{\_}err                & float64  & $e^-/s$       & Corrected PSF flux error. \\
5  & SAP{\_}Flux                & float64  & $e^-/s$       & SAP flux. \\
6  & SAP{\_}Flux{\_}err         & float64  & $e^-/s$       & SAP flux error. \\
7  & PSF{\_}flux{\_}nova        & float64  & $e^-/s$       & Mean PSF flux. \\
8  & PSF{\_}flux{\_}nova{\_}err & float64  & $e^-/s$       & Mean PSF flux error. \\
9  & SAP{\_}BKG                 & float64  & $e^-/s$       & SAP background flux. \\
10 & Centroid{\_}Column         & float64  & pix           & Centroid column value. \\
11 & Centroid{\_}Row            & float64  & pix           & Centroid row value. \\
12 & Red{\_}chi2                & float64  & -             & Reduced chi-squared value between PSF model and data. \\
13 & SAP{\_}quality             & int32    & -             & Quality flag from the TPF. \\
\enddata
\end{deluxetable*}

\restartappendixnumbering

\section{KBonus Source Catalog} \label{appx:catalog}

The catalog released with this work contains the list of extracted sources resulting in a light curve FITS file. It also contains extraction metrics and availability flags that can be used to filter sources. Table \ref{tab:cat} shows all the fields available in this catalog.

\begin{deluxetable*}{lllcl}[htb!]
\tablecaption{Description of Columns in Source Catalog.
\label{tab:cat}}
\tablewidth{0pt}
\tablehead{
\colhead{Column} & \colhead{Field} &\colhead{Format} &\colhead{Units} &\colhead{Description}
}
\startdata
1  & gaia{\_}designation        & String  & -       & Gaia designation number                           \\
2  & ra                         & Float32 & deg     & Right Ascension                                   \\
3  & dec                        & Float32 & deg     & Declination                                       \\
4  & sap{\_}mean{\_}flux        & Float32 & $e^-/s$ & Mean of SAP flux                                  \\
5  & sap{\_}mean{\_}flux{\_}err & Float32 & $e^-/s$ & Mean of SAP flux error                            \\      
6  & psf{\_}mean{\_}flux        & Float32 & $e^-/s$ & Mean of PSF flux                                  \\
7  & psf{\_}mean{\_}flux{\_}err & Float32 & $e^-/s$ & Mean of PSF flux error                            \\
8  & flfrcsap                   & Float32 & -       & Minimum detected SAP flux fraction                \\
9  & crowdsap                   & Float32 & -       & Minimum detected SAP crowding                     \\
10 & npixsap                    & Float32 & -       & Minimum detected SAP number of pixels             \\
11 & psffrac                    & Float32 & -       & Minimum detected PSF fraction                     \\
12 & pertrati                   & Float32 & -       & Mean detected perturbed/mean PSF ratio            \\
13 & pertstd                    & Float32 & -       & Minimum detected perturbed PSF standard deviation \\
14 & psf{\_}avail               & String  & -       & String encoding PSF flux availability per quarter \\
15 & sap{\_}avail               & String  & -       & String encoding SAP flux availability per quarter \\
16 & psf{\_}fraction{\_}flag    & String  & -       & String encoding PSF fraction quality per quarter  \\
17 & phot{\_}g{\_}mean{\_}mag   & Float32 & mag     & Gaia G band mean magnitude                        \\
18 & phot{\_}bp{\_}mean{\_}mag  & Float32 & mag     & Gaia BP band mean magnitude                       \\
19 & phot{\_}rp{\_}mean{\_}mag  & Float32 & mag     & Gaia RP band mean magnitude                       \\
20 & tpf{\_}org                 & Int32   & -       & TPF where sources was detected                    \\
21 & kic                        & Int32   & -       & Kepler input catalog number                       \\
22 & kic{\_}sep                 & Float32 & arcsec  & Distance between KIC and Gaia DR3                 \\
23 & kepmag                     & Float32 & mag     & Kepler magnitude                                  \\
24 & file{\_}name               & String  & -       & FITS file name                                    \\
\enddata
\end{deluxetable*}

\section{Data Bundles} \label{appx:bundles}

To facilitate data access to specific source types, we have created the following data bundles containing the light curves and the source catalog:

\begin{itemize}
    \item \textbf{M-dwarfs}: contains a total of $29,800$ sources. We follow the object selection described in Section \ref{subsec:demographics}.
    \item \textbf{KOIs and neighbors}: it packages light curves listed in the NASA Exoplanet Archive, including confirmed and false positive candidates. It also includes the light curve of neighbor sources around each KOI in a $30 \arcsec$ radius. This data bundle is useful for users that desires to explore false positives candidates and their neighbors.
    \item \textbf{White dwarfs}: contains a total of $91$ light curves as described in Section \ref{subsec:demographics}.
\end{itemize}

The files are stored in the Mikulski Archive for Space Telescopes (MAST) archive \footnote{KBonus Kepler Background \dataset[10.17909/7jbr-w430]{\doi{10.17909/7jbr-w430}}} and can be downloaded in bulk mode.

\bibliography{bibtex.bib}


\end{document}